\documentclass[
aps,%
12pt,%
final,%
notitlepage,%
oneside,%
onecolumn,%
nobibnotes,%
nofootinbib,%
superscriptaddress,%
noshowpacs,%
centertags]%
{revtex4}

\begin{document}
\selectlanguage{english}

\title{Comparative Analysis of the Observational Properties
of Fast Radio Bursts at the Frequencies of 111 and 1400 MHz}

\author{\firstname{V.~A.}~\surname{Fedorova}}Fedorova V.A.
\email{fedorova@prao.ru}
\affiliation{Pushchino Radio Astronomy Observatory, Astro Space Center, Lebedev Physical Institute, Pushchino, Moscow region, Russia}
\author{\firstname{A.~E.}~\surname{Rodin}}Rodin A.E. 
\email{rodin@prao.ru}
\affiliation{Pushchino Radio Astronomy Observatory, Astro Space Center, Lebedev Physical Institute, Pushchino, Moscow region, Russia}


\begin{abstract}
A comparative analysis of the observational characteristics of fast radio bursts at the frequencies 111 and 1400 MHz is carried out. The distributions of radio bursts by the dispersion measure are constructed. At both frequencies, they are described by a lognormal distribution with the parameters $\mu =6.2$ $\sigma = 0.7$. The dependence $\tau_{sc}(DM)$ of the scattering value on the dispersion measure at 111 MHz and 1400 MHz is also
constructed. This dependence is fundamentally different from the dependence for pulsars. A comparative
analysis of the relationship between the scattering of pulses and the dispersion measure at 1400 MHz and
111 MHz showed that for both frequencies it has the form $\tau_{sc}(DM)\sim DM^k$, where $k = 0.49 \pm 0.18$ and $k = 0.43 \pm 0.15$ for the frequencies 111 and 1400 MHz, respectively. The obtained dependence is explained
within the framework of the assumption of the extragalactic occurrence of fast radio bursts and an almost uniform distribution of matter in intergalactic space. From the dependence $\tau_{sc}(DM)$  a total estimate of the contribution to the matter of the halo of our and the host galaxy to $DM$ is obtained $DM_{halo} + \frac{DM_{host}}{1+z}\approx$\thickspace60\thickspace pc/cm$^3$. Based on the Log $N$ -- Log $S$ dependence, the average spectral index of radio bursts is derived $\alpha = - 0.63 \pm 0.20$ provided that the statistical properties of these samples at 111 and 1400 MHz are the same.

\end{abstract}

\maketitle

\section{INTRODUCTION}

One of the most interesting areas of modern astrophysics is the study of fast radio bursts (FRB), which
is a mysterious phenomenon that explains the nature
of which catastrophic cosmic events are involved,
from mergers of black holes and neutron stars to magnetar flares and asteroid vaporization by a flow of
charged particles in the vicinity of pulsars \cite{Fuller, Katz, Dai, Margalit, Istomin}. The recent registration of a powerful pulse from the well known magnetar SGR 1935+2154  \cite{Scholz} with a peak flux
density of the order of MegaJansky significantly
strengthened the position of supporters of magnetars
as sources of FRBs. Nevertheless, in our opinion, this
event does not completely cancel out other mechanisms of the occurrence of FRBs. 

Until recently, FRBs were recorded sporadically
either in archival data or during observations of other
space objects, and statistics on them was extremely
scarce. The situation has changed with the introduction of special monitoring radio telescopes operating
continuously around the clock. Abroad, such tools
include, i.e., CHIME \cite{CHIME}, ASKAP \cite{ASKAP}. There is also
such a radio telescope in Russia the Large Phased
Array (LPA) of the Pushchino Radio Astronomy
Observatory of the ASC FIAN (LPA FIAN).

The round-the-clock operation mode and the
completeness of the received data of the LPA FIAN
almost immediately after the launch of the monitoring
mode in 2012 led to the idea of trying to register pulse
signals of cosmic origin. This idea was first implemented to search for new pulsars \cite{Rodin}. In the fall of
2017, the development of an algorithm for detecting
single pulse signals was started. The first results were
published in the article by \cite{Fedorova1}. Further work was developed in the search for FRBs in the direction of
two nearby galaxies: M31 and M33. Nine more single
pulses were detected, including one repeating pulse.
All the events showed no apparent concentration
towards the centers of the M31 and M33 galaxies \cite{Fedorova2}. In order to start making meaningful reliable conclusions about the properties of new pulses, it was necessary to increase their statistics to several dozen, so a
search was started for the entire available area of the
sky. At the time of writing this article, 63 pulses were
detected with the FIAN LPA telescope. Although the
authors defined them for themselves from the very beginning as FRBs, it was necessary to study their
observational properties and compare them with similar properties of radio bursts observed at other frequencies in order to speak confidently about this. This
is the main purpose of writing this article.

Furthermore, we will carry out a comparative analysis of such observational characteristics as the dispersion measure, the dependence of the pulse scattering
on the dispersion measure, which, as will be shown
later, is fundamentally different from the similar
dependence for pulsars, and, finally, the log $N$ -- log $S$
dependence at the frequencies 111 and 1400\thinspace MHz,
from which the average spectral index of the pulses will
be derived.

The distribution density of the detected radio
bursts by the dispersion measure has been studied by
many authors. For example, in the work of Cordes
et al. \cite{Cordes16}, the integral distribution density of radio bursts is analyzed in comparison with the spatial distribution of free electrons and radio bursts, and it is
concluded that the observed distribution can be
explained by a model that includes a dense core and a
more rarefied halo. In the work of Dolag et al. \cite{Dolag}, the authors conclude that the observed $DM$ distribution is
consistent with the cosmological population at redshifts $z$ = 0.6 -- 0.9, regardless of how the FRBs are distributed with respect to the large-scale structure or
properties of the host galaxies.

The scattering of pulses by inhomogeneities of the
medium is an important tool for studying the properties of both the interstellar and intergalactic medium.
Pulsars are most suitable for the role of pulsed radiation sources for studying the properties of the interstellar medium. The scattering of radio waves by small scale fluctuations of electron density in the interstellar
medium was first recognized as the cause of changes in
the radiation intensity of pulsars in \cite{Scheuer,Rickett}. The galactic medium is very heterogeneous, and a detailed study
of the processes taking place inside it began almost
immediately after the discovery of pulsars. At that
time, it was possible to study the interstellar medium at
relatively small distances from the observer, since the
first recorded pulsars were close objects and had a relatively high flux densities \cite{hewish, Robinson, Alekseev}. Cordes et al. \cite{Cordes} were the first to propose a two-component model of
turbulence in the galactic medium. The first component is understood by the authors as an inhomogeneous medium in the region up to galactic heights $z$ < 100\thickspace pc, which is associated with the type I population of the Galaxy. The region at a distance of $\gtrsim$ 0.5\thickspace kpc is the second component of the model and
determines the scattering at high galactic latitudes $|b|\thickspace\gtrsim$ 10$^\circ$. On such scales, the medium, according to
the authors, is practically homogeneous. This result
turned out to be one of the most important, since it
was previously believed that turbulent fluxes are distributed non uniformly.

The question related to the mechanism of radiation
scattering was considered in more detail by Sutton \cite{Sutton}. It describes in detail all factors leading to multipath propagation of the signal, and also shows the
theoretical relationship between the pulse broadening,
the decorrelation frequency and the dispersion measure. The paper also considers anomalously strong
scattering at small dispersion measures. Such an effect
can be observed as a result of the interaction of the
pulse with strong turbulent streams, i.e., in the HII
regions. Also, an example of a strong dependence of
scattering on the dispersion measure is given in the
work of Bhat et al. \cite{Bhat}. The authors show that for
galactic pulsars with the same values of $DM$, the scattering value can differ by three orders of magnitude.

The main contribution to the scattering of pulses is
made by the galactic medium, where the density of
matter is significantly higher compared to the intergalactic medium. Accordingly, the scattering of pulses in
the galactic medium should depend more strongly on
the dispersion measure, which cannot be stated
regarding the pulses scattered by the intergalactic
medium. This effect was indicated in the works of
Lorimer and Karastergiou  \cite{Lorimer, Karastergiou}. It follows from
their studies that the scattering of FRB in the intergalactic medium is extremely small or even completely
absent, and that pulses recorded at frequencies below
1 GHz experience much less broadening, unlike pulses
of galactic origin. According to the authors, this makes
it possible to observe FRBs at lower frequencies.

In the work of Zhu et al. \cite{Zhu}, the authors consider
the scattering of FRBs in the intergalactic medium at
large cosmological distances, since the average distance to the pulse origin region is 400-500 Mpc. In the
article, it is shown by modeling that the scattering of
pulses in the voids is rather weak, but can be amplified
by the inhomogeneous distribution of gas in galactic
clusters and filaments. In the article \cite{Dolag}, Dolag and
co-authors model the distribution of intergalactic
matter depending on the large-scale structure of the
Universe. They also estimate the contribution of the
matter of our Galaxy to be approximately two times
higher than it was before them, due to the contribution
of the Galactic halo at a level of $DM_{halo} \thickspace \approx$\thickspace30\thickspace pc/cm$^3$.

In contrast to the above-mentioned articles \cite{Zhu} and \cite{Dolag} , although we assume that the intergalactic
matter is inhomogeneous, the contribution of intergalactic matter scattering can be neglected and considered homogeneous in the first approximation in comparison with the contribution to the scattering of
impulses from the matter of our galaxy and possibly
the host galaxy.

Returning to the question of scattering at low frequencies, we can cite the work of Kuz'min et al. 2007 \cite{Kuzmin2007} as an example. It measured the broadening of
pulses from a sample of 100 galactic pulsars at frequencies of 102 and 111 MHz. As a result of the analysis, it was found that the dependence of the scattering
value on the dispersion measure is described by the
power law:

\begin{equation}
\tau_{sc}(DM) = 0.06\left(\frac{DM}{100}\right)^{2.2\pm 0.1} [\rm c]
\end{equation}

Kuz'min et al. also showed in \cite{Kuzmin2007} that, at a distance of
up to 3 kpc, the turbulent fluxes of the scattering
medium are statistically homogeneous. But this conclusion was made for the galactic regions. Our work is
aimed at studying the effects that occur in the intergalactic environment. Pulsed sources such as radio
bursts, which are currently the only means for studying
intergalactic propagation effects, are fully suitable for
these purposes.

In this work, the scattering value of 63 FRB
recorded at a frequency of 111 MHz was studied \cite{Fedorova1, Fedorova2}. The dependence of scattering $\tau_{sc}$ on the dispersion
measure $DM$ from Kuzmin's article \cite{Kuzmin2007} is constructed for pulsars with a dispersion measure from 2.97 to 196\thickspace pc/cm$^3$. It was decided to supplement this
dependence based on measurements of FRB at
111 MHz, the dispersion measure of which is in the
range from 172\thickspace pc/cm$^3$ to 1868\thickspace pc/cm$^3$.

The integral distribution of FRB by the peak flux (Log $N$ -- Log $S$) was analyzed in the works of
Oppermann, Macquart and Popov \cite{Oppermann, Macquart, Popov}.  To consider the dependence, pulses registered at frequencies > 700\thickspace MHz were used. Oppermann et al.  \cite{Oppermann} show
that the distribution of pulses by the flux is consistent
with a uniform distribution of sources in Euclidean
space. In \cite{Macquart}, Macquart and Ekers show that the distribution of FRBs in space not only corresponds to the
law $S^{-3/2}$,  but may also have a steeper dependence. In
the article \cite{Popov}, the integral fluence distribution $N(>F) - F$ is analyzed and it is shown that the pulses
are divided into two populations: 0.5 < $F$ < 3 Jy $\cdot$ ms and 3 < $F$ < 100 Jy $\cdot$ ms. This feature, according to the
authors, can be explained either by small statistics or
by the effect of selection.

In the following sections, the technical characteristics of the LPA FIAN radio telescope are given, the
method for measuring the pulse width of FRB is
described, and the results obtained are discussed.

\section{EQUIPMENT}

The observations were carried out with the Large
Phased Array (LPA LPI). This is a meridian-type
instrument, in which the field of view shows a part of
the sky falling from +42.13 to --8.20 degrees in declination. The antenna has a large field of view, which
is $\sim$50\thickspace sq. deg. The effective area of the LPA LPI is 47\thickspace000 m$^2$ and has a maximum value at the zenith.
This value decreases to the horizon proportionally to cos\thickspace$z$, where $z$ -- is the zenith distance. The directional
beam (DB) of the LPA LPI is of particular interest. It includes a declination-controlled directional beam
(DB-1) and a stationary directional beam (DB-3).
Pulsars are observed and studied with DB-1. Round-the-clock monitoring of various sources is carried out
with DB-3.

The radio telescope receives radiation in the frequency range 111\thickspace MHz $\pm$ 1.25\thickspace MHz. The recording is
carried out in two modes using a multi-channel digital
receiver. The first mode includes observations with
low frequency resolution in six frequency channels of
415 kHz each. In this case, the time resolution is 0.1 s.
In the second mode, the recording is carried out in
32 frequency channels, each of which is 78 kHz, and
the time resolution is 12.5 ms. Both recording modes
are obtained by converting the signal on the FFT processor from 512 channels \cite{Oreshko}.

The fluctuation sensitivity of the radio telescope in
the low-resolution mode is 140 mJy, which makes the
instrument one of the best in the world. The self-noise
temperature in the system is in the range from 550 to
3500 K and depends on the sky background.

During the observations at the LPA LPI for the
period from 2012 to the present, a huge amount of data
has been accumulated. As a result of careful processing
of data obtained over eight years in daily observations
of several sections of the sky with a total area of
310 square degrees, more than sixty new radio bursts
were detected. Currently, the search continues
throughout the entire area of the sky available for
observation.

\section{CATALOG "PRAO FRBs at 111 MHz"}

In the work of Fedorova and Rodin \cite{Fedorova2}  it was shown that the
average rate of registration of bursts by the LPA LPI radio telescope at a frequency of 111 MHz is $\sim$ 2000  pulses/year. At the beginning of 2021, the
archival data of 2018 was processed from January to
June in the sky area $\alpha=11^h45^m \div 12^h45^m$ and $\delta=21.38^{\circ} \div 41.72^{\circ}$. During this period, 51 new phenomena were detected, which, in terms of the entire sky for
the year, corresponds to the number of 10$^3$ pulses/day
and corresponds to the estimates given in various studies \cite{PetroffBIG}. Due to the fact that the number of detected
pulses exceeded several dozen, a separate catalog of
radio bursts registered at 111 MHz was created -- "PRAO FRBs at 111 MHz"\footnote{https://www.frb.su/catalogue-prao-frb}.This resource is freely
available and includes the following parameters:

\begin{enumerate}
\catcode`\_=\active
\item FRB_name -- the name of the pulse. It is given in
the form of FRB yymmdd.Jra+dec \cite{CHIME}.
\item RA, DEC --the pulse coordinates are given in the
format RA(J2000) hh:mm:ss, DEC (J2000) deg. The
estimate of the error in determining the right ascension is $\pm2^m$, error of the declination corresponds
to $\pm15'$.
\item GL, GB -- the galactic coordinates for the epoch
J2000.
\item Date -- the date in the yyyy/mm/dd format corresponds to the date in the archive data of the moment
when the pulse was detected.
\item UTC_time -- the UTC value in the format
hh:mm:ss.
\item Flux -- the value of the peak flux of FRB (Jy).
The accuracy estimate is $\pm0.05$ Jy.
\item DM -- the value of the dispersion measure (pc/cm$^3$). The error estimate is $\pm5$\thickspace pc/cm$^3$.
\item S/N -- the value of the signal-to-noise ratio. For
the previously detected impulses from \cite{Fedorova1, Fedorova2}, the
S/N value was recalculated to a more conservative
side.
\item Width_observed -- the width of the total pulse
after convolution with the template (ms). The measurement error is $\pm100$ ms.
\item Width_original -- the initial width of the scattered pulse received by the antenna (ms). The error is $\pm100$ ms.
\item Fluence -- the energy density of a fast radio burst
(Jy $\cdot$ ms).
\item z_YMW16 -- the value of the redshif $z$ in accordance with the YMW16 electron-density model \cite{YMW16}. This value is only an estimate, since it depends on the
accepted value of the average density in the intergalactic medium.
\end{enumerate}

The completeness of the catalog was evaluated by
analyzing the difference between the histogram of the
distribution by $DM$ and the lognormal distribution (see Figure \ref{ris:fig7} below). The histogram shows that there
is a shortage of pulses at $DM < 200$ and $DM > 700$. The deficit is a fraction $\sim 0.32$ of the total distribution
area. This value is close to the proportion of discarded
pulses $\sim 43\%$ that were rejected due to strict selection
criteria: if the pulse was not visible in all six channels,
then it was not included in the catalog, although it was
most likely a real pulse. Thus, according to our estimates, the catalog of radio bursts at 111 MHz is full at $\sim 60-70\%$.

Below, for each pulse, there is a dynamic spectrum
in six frequency channels (upper image) and a profile
(lower image). The pulse amplitude is given in ADC
units, the countdown is understood as a unit of time
equal to 0.1 s.

\section{DATA PROCESSING}

The method of detecting single pulse signals is
described in detail in previous articles by Fedorova and
Rodin \cite{Fedorova1, Fedorova2}, so we will focus on it briefly. Since the
LPA antenna receives a signal scattered on the inhomogeneities of the cosmic plasma, and the signal
undergoes additional broadening when received in the
frequency band, a convolution with a template of a consistent shape (an exponentially decaying pulse) is
used to isolate it. In radiophysics, this approach is
called a correlation receiver, and the result of convolution of a signal with a template is a signal function.
Since the typical values of the dispersion measures of
the detected FRB are several hundred pc/cm$^3$, it was
decided to use a template with a characteristic width $t_s=1$ s, which according to formula (1) corresponds to
the dispersion measure $DM=360$ pc/cm$^3$. As was
shown in the work of the authors [10], such a correlation approach can significantly improve the signal-to-noise ratio of the detected signal, as well as improve
the level of fluctuation sensitivity up to 44 mJy. The
article \cite{Fedorova2}  shows an example of recording before and
after using a convolution with a template, where it is
clearly visible that it is not possible to isolate a pulse
signal without using a correlation receiver.
 
\begin{figure}[h!]
	\setcaptionmargin{1.3mm}
	\vbox{\includegraphics[width=1.05\linewidth]{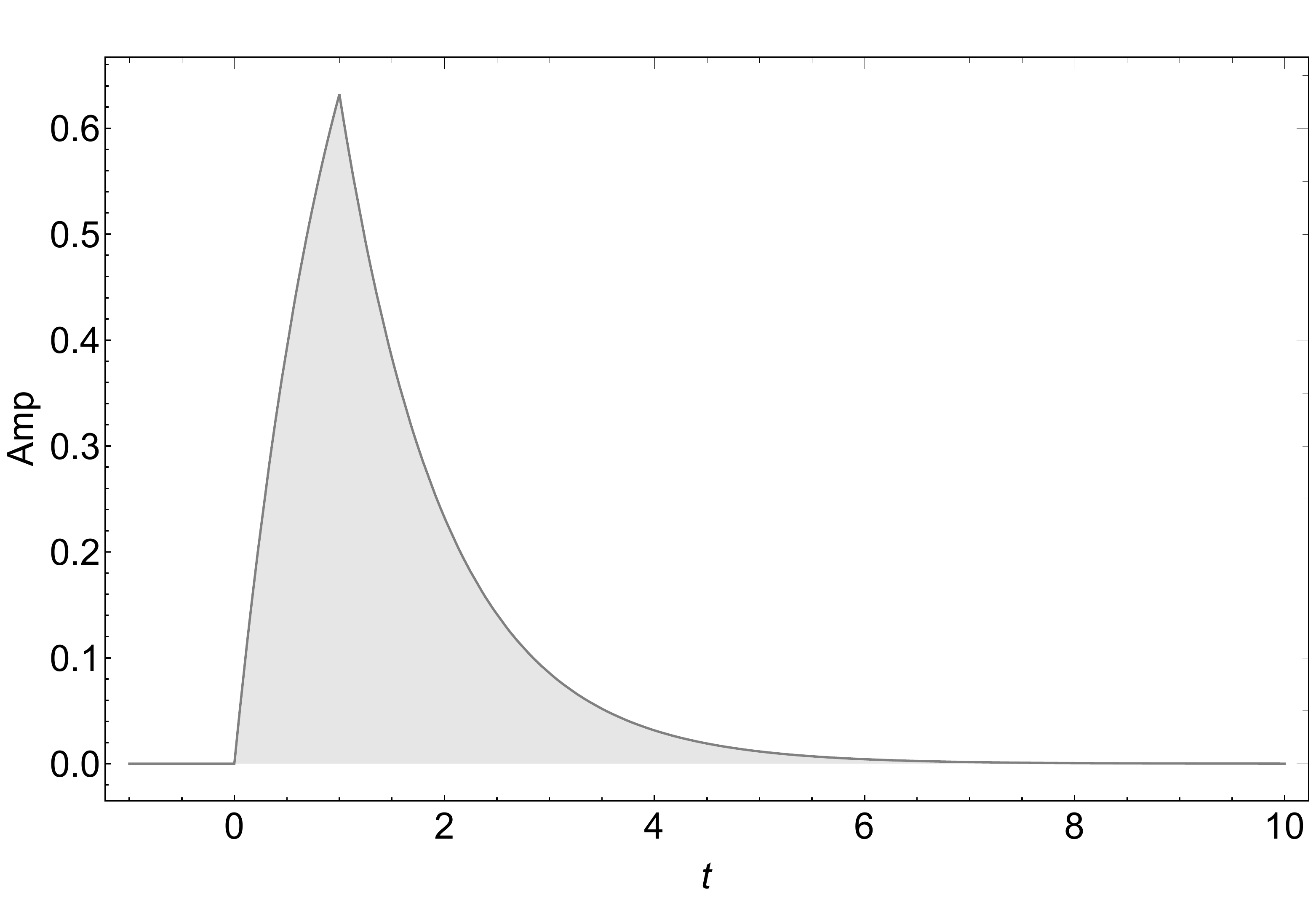}}
	\caption{A template similar to the shape of the received pulse. Amp is the signal amplitude, is the time, s.}
	\label{ris:fig1}
\end{figure}

In this article, we consider the broadening of the
pulse due to scattering, so we will further describe the
procedure for measuring the width of the radio burst.
Analytically, the shape of the template $p(t)$ and the
scattered pulse $s(t)$ is written as:
\begin{equation}
(p(t), s(t))=(a,b)\left\{
\begin{array}{ll}
1-\exp(-t/t_s), & 0\leq t\leq \tau,\\
\exp(-t/t_s)(\exp(-t/t_s)-1), & 0\leq \tau \leq t,
\end{array}\right.
\end{equation}
where $a,b$ -- are the amplitudes of the template and the
pulse, respectively, $t_s$ -- is the scattering of the pulse, $\tau$ -- is
the broadening of the pulse in the frequency channel.
For the template, as already mentioned earlier, the
value $t_s=1$ s. It is also necessary to require the fulfillment of the condition $\int_0^\infty p(t)=1$ for the conservation
of the pulse energy. The shapes of template and pulse
are shown in Fig. \ref{ris:fig1}.
 
\begin{figure}[h!]
	\setcaptionmargin{1.3mm}
	\vbox{\includegraphics[width=1.05\linewidth]{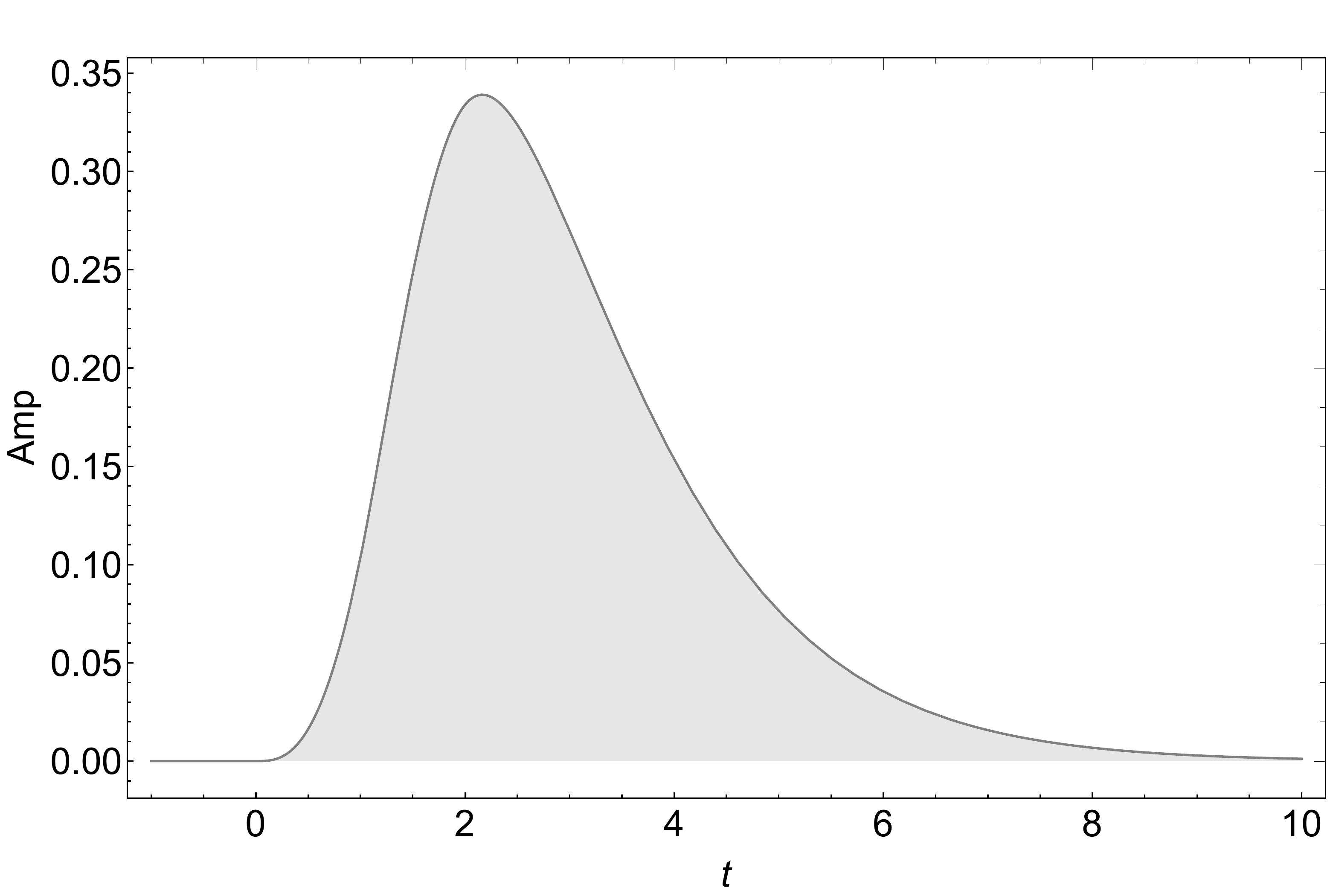}}
	\caption{ The signal function $f(t)$ which is the result of convolution of a noisy signal $s(t)$ and a template $p(t)$. Amp is the signal
	amplitude, $t$ is the time.}
	\label{ris:fig2}
\end{figure}

The signal function is a convolution of the noisy
signal $s(t)$ and the template $p(t)$:
\begin{equation}
f(t)=\int_{-\infty}^{\infty}p(t)s(t - t_1) dt_1.
\end{equation}
Its graph is shown in Fig. \ref{ris:fig2}.
 
Since the shape of the signal function is strongly
distorted by the influence of noise, the practical
approach used Gaussian insertion to measure the
pulse width, and the asymmetry of the pulse was not
considered. The position of the pulse was determined
by the position of the maximum in the recording, the
pulse amplitude was reduced to one. Thus, the only
parameter to be determined was the pulse width $\sigma$. Since the Gaussian parameter $\sigma$ is determined at the
height $1/\sqrt{e}$, and the scattering value is measured at the
height $1/e$, the scattering of the pulse received by the antenna was calculated using the formula $\tau_{sc}=2\sqrt{2}\sigma-\tau-t_s$, where $\tau$ -- is the broadening in one frequency channel, and $t_s=1$ s is the characteristic width
used for smoothing the template.

\section{RESULTS}

The results of measurements of the scattering value
of 63 pulses, as well as some parameters of FRB from
the catalog "PRAO FRBs at 111 MHz", are presented in Table 1. It contains the names of twelve previously
detected pulses \cite{Fedorova1, Fedorova2} and 51 new ones. If the pulses
were recorded on the same day, but in different beams
of the radiation pattern of the LPA FIAN radio telescope, then the coordinates were added to the standard form of the name of the phenomenon. The coordinates of each pulse for the epoch J2000 are given in
the second column of Table 1. The third column contains the dispersion measure of radio bursts, measured
with an accuracy of $\pm5$\thickspace pc/cm$^{3}$.  The last column of the
table contains the value of the pulse scattering $\tau_{sc}$.

A comparison of the two $DM$ distributions at
111 and 1400 MHz showed that both of them are
described by the lognormal distribution (4) with the parameters $\mu = 6.1 \div 6.2$, $\sigma = 0.7$ and, thus,
at a statistically significant level, they coincide and correspond to the range of characteristic values $DM$\thickspace=\thickspace252\thickspace$\div\thinspace954$ pc/cm$^{3}$:

\begin{equation}
P(x) = \frac{e^{-\frac{(\mu + {\rm ln} x)^2}{2 \sigma^2}}}{x\sqrt{2 \pi} \sigma}
\end{equation}

In some papers, the authors derive
distributions by subtracting the contribution of the
Galaxy $DM_{MW}$ from the total $DM$. In this paper, we
use the full value of $DM$,  since $DM_{MW}$ is often known
with a relative accuracy of 0.5 and, thus, its exclusion
introduces an additional error in the distribution.

\begin{figure}[h!]
	\setcaptionmargin{1.3mm}
	\vbox{\includegraphics[width=1\linewidth]{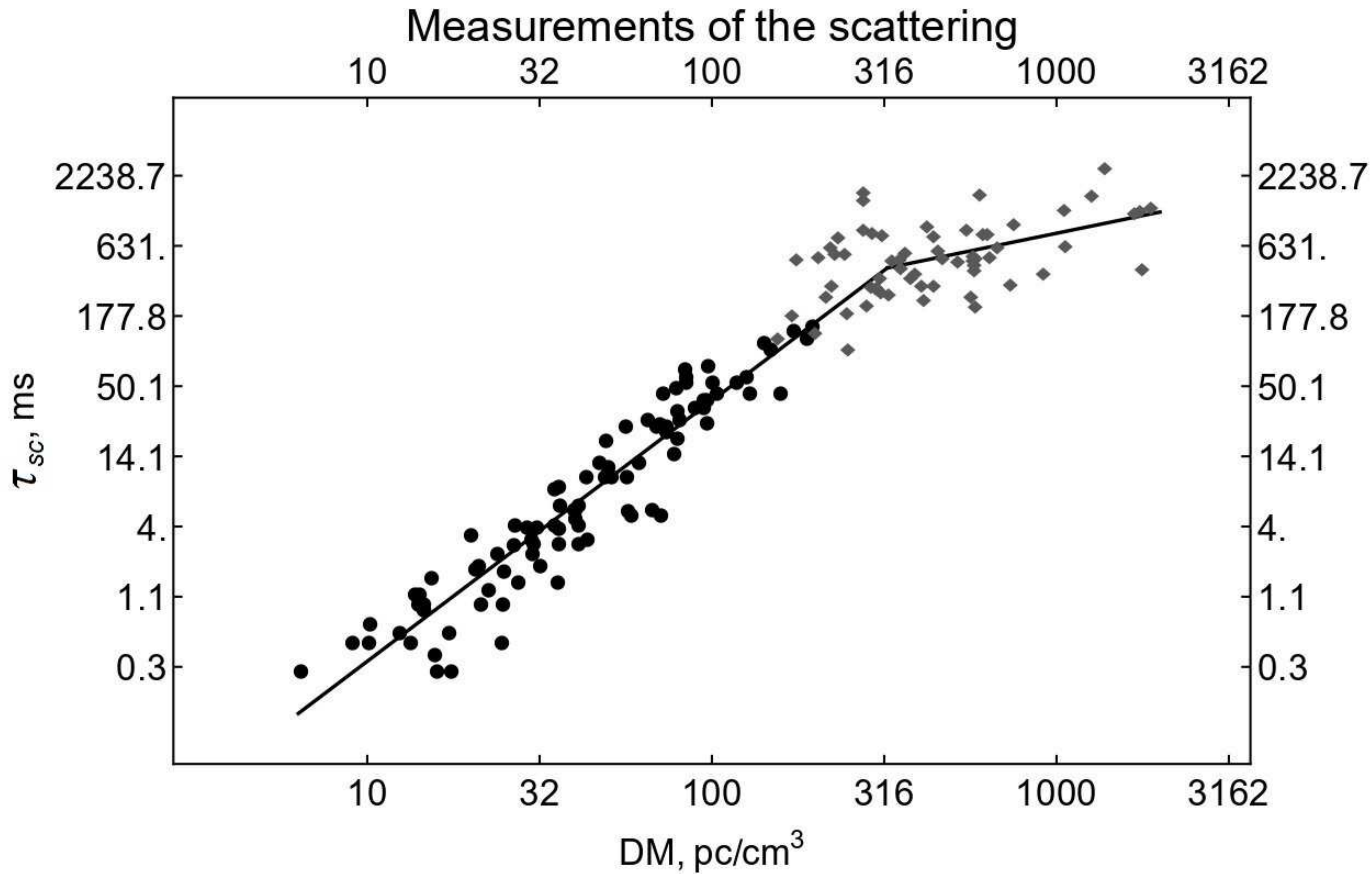}}
	\caption{Dependence of the broadening of pulsar pulses and radio bursts due to scattering on the dispersion measure $\tau_{sc}(DM)$. The
	black circles show measurements at 111 MHz for pulsars. The gray diamonds show measurements at 111 MHz for fast radio bursts.}
	\label{ris:fig3}
\end{figure}

Based on the data of Table 1, a graph of the dependence of the scattering value $\tau_{sc}$ on the dispersion
measure $DM$ was constructed together with the
dependence for pulsars given in the article \cite{Kuzmin2007}. This
graph is shown in Fig. \ref{ris:fig3}.

For FRBs, the power dependence is weaker in
comparison with the dependence for pulsars: if for
pulsars observed at a frequency of 111 MHz, the
authors \cite{Kuzmin2007} determine the slope coefficient $k = 2.2\pm0.1$ (formula (1)), then for FRBs recorded
also at a frequency of 111 MHz, we give the formula $\tau_{sc}(DM)=20.2\, DM^{0.49\pm0.18}$ ms.

\begin{figure}[h!]
	\setcaptionmargin{1.3mm}
	\vbox{\includegraphics[width=1\linewidth]{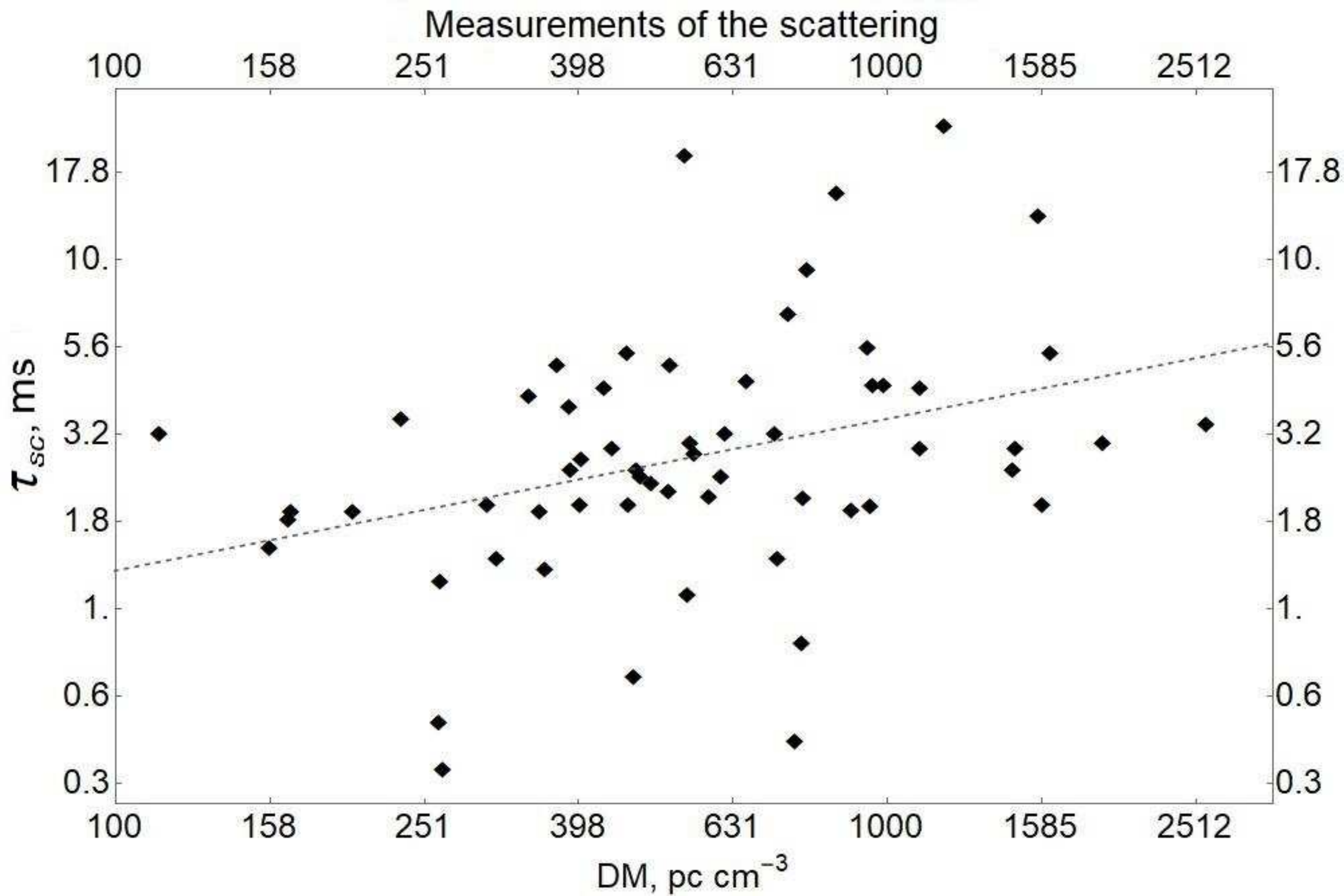}}
	\caption{Dependence of the broadening of radio burst pulses due to scattering on the dispersion measure at 1.4 GHz according to
	the FRB catalog.}
	\label{ris:fig4}
\end{figure}

The dependence $\tau_{sc}(DM)$ was also constructed for
FRBs at a frequency of 1.4 GHz from the FRB catalog \cite{frbcat}. The graph of this dependence is shown in Fig. \ref{ris:fig4}. To construct the dependence, 59 pulses with a scattering value from 0.34 ms to 24.3 ms and a dispersion
measure from 114 pc/cm$^{3}$ to 2596\thickspace pc/cm$^{3}$. were used.
In this case, the dependence of scattering on $DM$ is
described by the formula $\tau_{sc}=0.176\, DM^{0.43\pm 0.15}$ ms.
The slope coefficient $k = 0.43\pm0.15$ within the error
is consistent with the result obtained for radio bursts at
111 MHz.

\begin{figure}[h!]
\setcaptionmargin{1.3mm}
	\vbox{\includegraphics[width=1\linewidth]{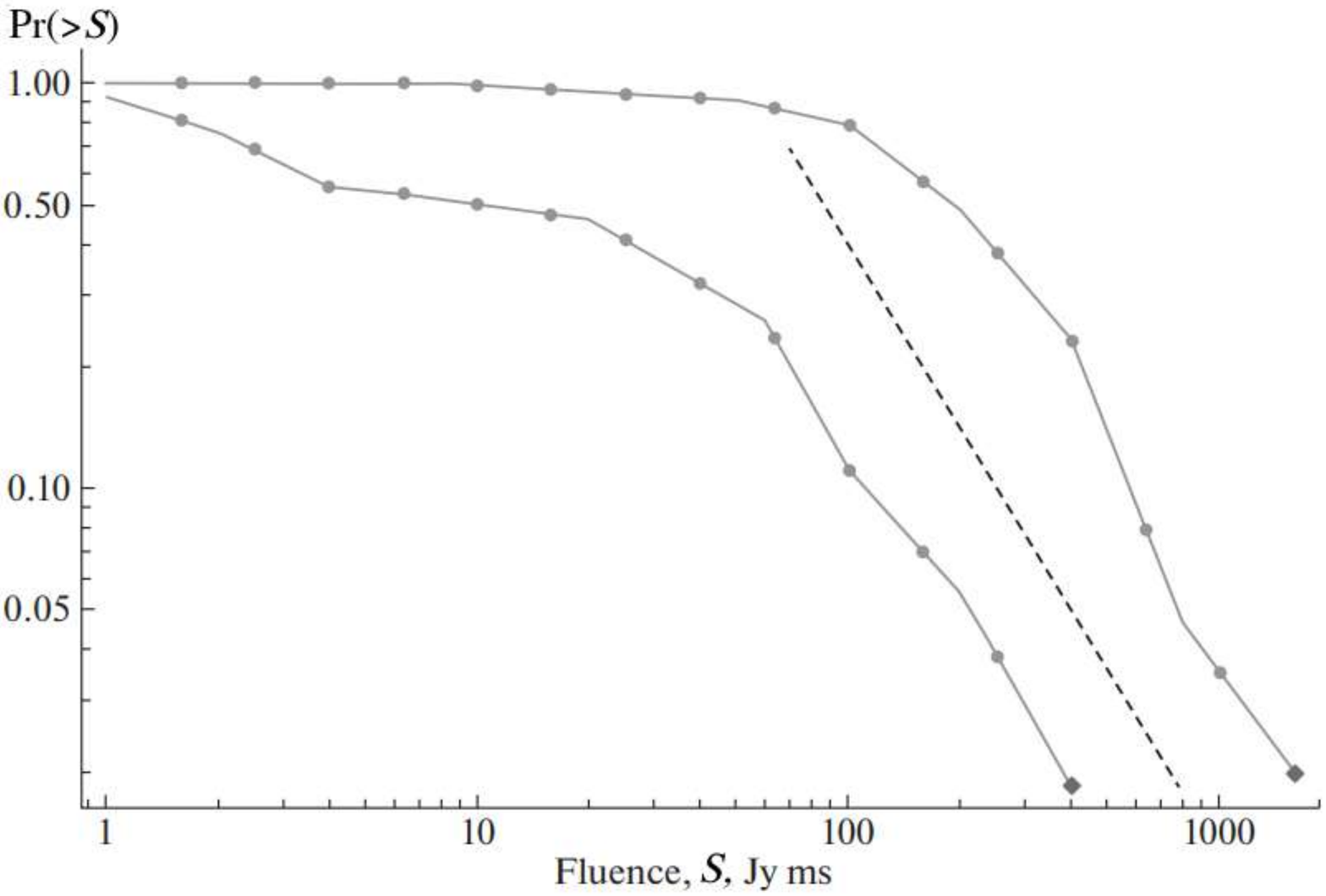}}
	\caption{Log$N$ -- Log$S$ dependence for fast radio bursts at frequencies of 111 MHz and 1.4 GHz. The lower curve corresponds to
	pulses registered at 1.4 GHz, the upper curve corresponds to pulses at 111 MHz. The dotted line between the curves corresponds
	to the slope -- 3/2.}
	\label{ris:fig5}
\end{figure}

Figure \ref{ris:fig5} shows a graph of the Log$N$ -- Log$S$, dependence, built at different frequencies. If we
assume that the observed samples have the same properties, then the average spectral index of radio bursts $\alpha$ = -- 0.63 $\pm$ 0.20 can be derived from the mutual shift
of the graphs. 

By analogy with pulsars, a low-frequency break
caused by absorption on free electrons or radiation
features should be observed in the spectra of radio
bursts, then it is possible to calculate the break frequency $f_1$ assuming several spectral indices at a frequency of 1400 MHz and a flat spectrum in the low frequency region, which is shown in Table 2.

\section{DISCUSSION}

The obtained power dependence $\tau_{sc}(DM)$ at a frequency of 111 MHz turned out to be weaker in comparison with $\tau_{sc}(DM)$ for pulsars (the exponent $k \sim 0.5$ ¢¬¥áâ® $k \sim 2$). The exponent $k$ within the
error is consistent with the exponent $k$ of the dependence $\tau_{sc}(DM)$ constructed for a sample of FRB at a
frequency of 1.4 GHz from the FRB catalog \cite{frbcat}.

Other authors have already considered the scattering of FRB on inhomogeneities of the intergalactic
plasma. For example, we can cite the work by Zhu
et al. \cite{Zhu2}, where the authors modeled scattering by
setting different sizes of inhomogeneities along the
line of sight. They obtained a dependence $\tau_{IGM}\sim {DM}_{IGM}^{1.6-2.1}$ that does not explain the experimental power law $\tau \sim DM^{0.5}$, although it fits into the
spread of data on the graph $\tau$ -- $DM$.
 
In our work, we explain the obtained relationship
within the framework of the idea in which the pulse is
scattered along the entire line of sight, but the main
scattering occurs in the host and our galaxy, and the
intergalactic medium on the propagation path is
sparse and does not have a dominant influence on the magnitude of the scattering of pulses despite large values of $DM$. Once again, we emphasize that we do not
cancel scattering in the intergalactic environment, but
neglect it. The magnitude of the exponent $k$ in the
experimental dependence $\tau_{sc}(DM)$ is determined by
the position of the scattering screen along the "source--observer" line. We associate the position of the scattering screen with the boundary of our galaxy or the
host galaxy, where the direction of propagation of the
pulse deviates from the initial one. A classic example
of the model of scattering of a pulse signal on a thin
screen can be found in the article by Scheuer \cite{Scheuer}. With this approach, since we are dealing with a substance on the line of sight in the study of dependence $\tau_{sc}(DM)$ it is convenient to measure the distance in
units of $DM=\int\limits_{0}^{L}n_edl$. Obviously, since the concentration of matter in the galactic and intergalactic
medium differs significantly, the dependence $DM(L)$ will not be linear.

We divide the measured dispersion measure $DM$ of
the radio burst into the following components:

\begin{equation}
DM = DM_{MW} + DM_{halo} + DM_{EG} + \frac {DM_{host}}{1+z},
\end{equation}
where $DM_{MW}$ is the contribution of the Galaxy's matter, which is modeled based on observations of pulsars. The remaining amount $DM_{halo} + DM_{EG} + \frac{DM_{host}}{1+z}$ is
the contribution $DM_{halo}$ of the matter of the halo of
our Galaxy, $DM_{EG}$ is the contribution of extragalactic
matter, $\frac{DM_{host}}{1+z}$ is the contribution of the matter of the
host galaxy, and $z$ is its redshift. Of all these quantities, $DM$ and $DM_{MW}$ are well known. For the halo, as
mentioned earlier, an estimate of $DM_{halo} \approx$\thickspace30\thickspace pc/cm$^3$ is known.

\begin{figure}[h!]
	\setcaptionmargin{1.3mm}
	\vbox{\includegraphics[width=1\linewidth]{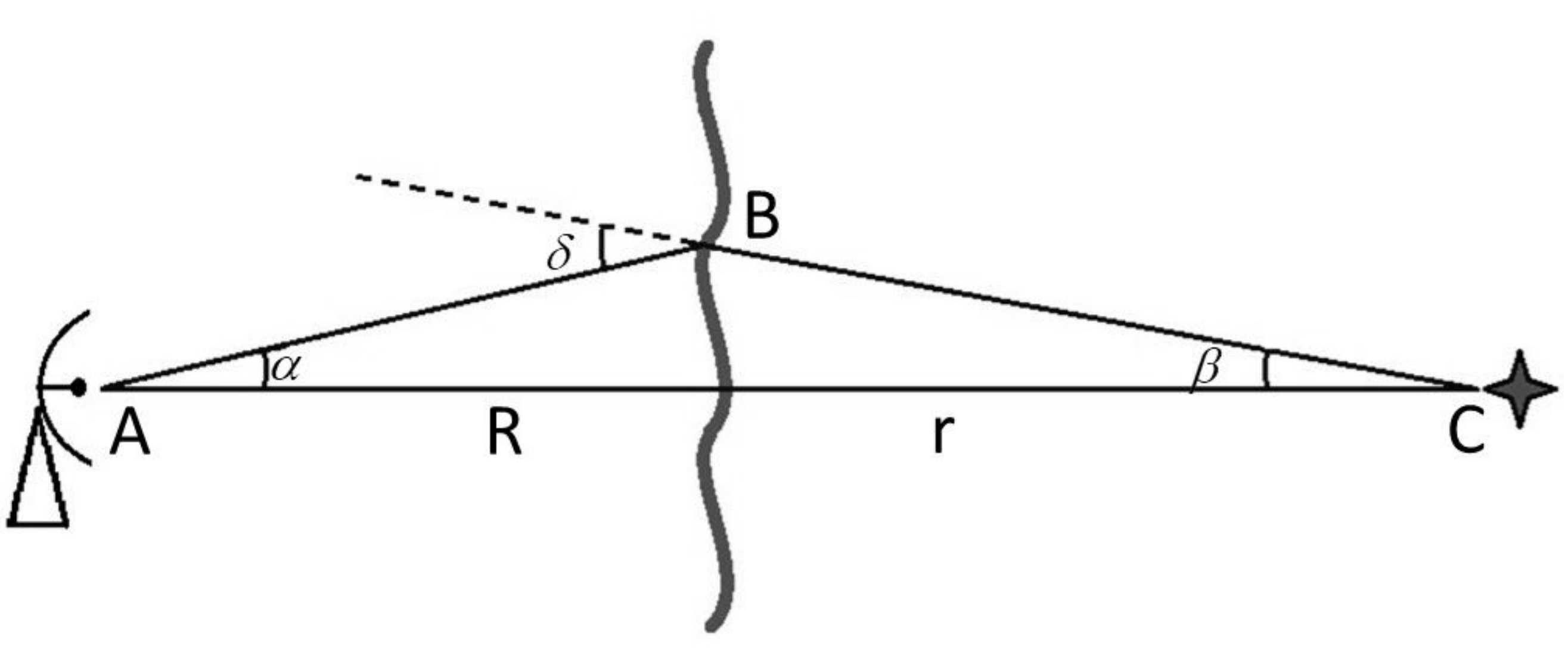}}
	\caption{The scheme of signal propagation from a source to an observer. The point $A$ is the observer, the point $C$ is the place of
	origin of the pulse, $B$ is the scattering screen, $R$ and $r$ are the distances from the observer to the screen and from the screen to the
	place of occurrence of a fast radio burst, and $\delta$ is the angle of deviation of the pulse.}
	\label{ris:fig6}
\end{figure}

Figure \ref{ris:fig6} shows the scheme of signal propagation
from the source to the observer. The point $C$ indicates
the area of occurrence of a fast radio burst, and the
point $A$ is the observer. The curved line in the diagram shows a thin scattering screen. $R$ and $r$ are the distance
from the observer to the scattering screen and from the
screen to the point of origin of the pulse, respectively.

From the point $C$,  the pulse propagates towards the
observer, then falls on the scattering screen, which
deflects the pulse by an angle $\delta$, and falls to the observer. In this case, the phase incursion $\Delta$, which we
associate with the scattering $\tau_{sc}$ and the shape of the
pulse, is defined as:

\begin{equation}
\Delta = AB + BC - AC = K
\frac{\delta^2}{2}R\left(1+\frac{R}{r+R}\right),
\end{equation} 
where $\delta$ = $\alpha$+$\beta$. In accordance with the use of quantities $DM$ as a characteristic of distance in formula (5),
the observed $DM\thickspace\equiv\thickspace r + R$ is responsible for the distance $r+R$, $DM_{MW} + DM_{halo} + \frac{DM_{host}}{1+z} \equiv R$ is
responsible for $R$, and $DM_{MW} \equiv r$ is responsible for $r$. Expression (6) is symmetric with respect to the variables  $r$ and $R$, Therefore, for simplicity, one screen
can be considered during modeling.

As part of the work, simulation was carried out,
which, as mentioned above, consisted in changing the
position of the screen along the line of sight. The distance $r + R$ to radio bursts was set in accordance with the experimental distribution $DM$ of all registered
radio bursts, which is shown in Fig.\ref{ris:fig7}.

The deviation angle $\delta$ was set as a normally distributed value with an average of 0 and $\sigma$ = $10^{-9}$ rad, which
corresponds in order to the observational data of pulsar scattering. The value $K$ was selected experimentally so that the scattering value $\tau_{sc}$ at the given $DM$ and $\delta$ corresponds to the observed one. The distance $R$ was determined by a lognormal distribution with
parameters $\mu = 2.7$ and $\sigma = 0.3$ corresponding to the
model distribution $DM_{MW}$ in our Galaxy \cite{YMW16, CordLaz} taken from the FRB catalog. The simulation results
are shown in Figure \ref{ris:fig8}. The experimental coefficient $k \sim 0.5$ of the dependence $\tau_{sc} \sim DM^k$ corresponds to the contribution to $DM$ of the host and our Galaxy at
the level of $DM_{MW} + DM_{halo} + \frac{DM_{host}}{1+z} $\thickspace=\thickspace105\thickspace pc/cm$^{3}$. For the characteristic value of $DM_{MW}\sim 45$\thickspace pc/cm$^{3}$, corresponding to the maximum of the lognormal distribution, we obtain that the total contribution of the
matter of the host and the halo of our Galaxy to the dispersion measure is $DM_{halo} + \frac{DM_{host}}{1+z} \sim 60$\thickspace pc/cm$^{3}$, which corresponds to the currently accepted value of $DM_{halo} \sim 30$\thickspace pc/cm$^{3}$ \cite{Dolag},  provided that the contributions of the matter of halo of our Galaxy and the matter of the host galaxy are equal.

It should be specially noted that a small number of
scattering screens $\leq$ long the path of propagation of
FRB is independently indicated by the shape of the
pulses observed at a frequency of 1.4 GHz, which is
close to the decaying exponent for the overwhelming
number of radio bursts. Specially conducted mathematical modeling clearly shows how the shape of the pulse changes when passing an increasing number of
scattering screens. In full accordance with the central
limit theorem, the shape of the pulse tends to the
Gaussian with an increase in the number of screens, as
shown in Fig. \ref{ris:fig9}.

Analysis of Log $N$ -- Log $S$ dependencies at two frequencies shows that, in general, these dependencies correspond to each other and none of them follows
exactly the law $S^{- 3/2}$. If we assume that the observed
pulse samples are statistically equivalent, then we
can deduce their average spectral index, which is
equal to $\alpha = -0.63 \pm 0.20$.  This value generally corresponds to the expected negative value, although a
number of authors prefer to give steeper indices $\alpha$\thickspace =\thickspace--1.8\thickspace $\div$\thickspace--1.5.

\section{CONCLUSIONS}

Attempts to detect radio bursts at low frequencies
have been made repeatedly by many researchers
abroad. These observations can be briefly described as
unsuccessful, since no radio bursts were detected at
frequencies below 300 MHz. This is an occasion for
the authors \cite{Pilia} to comment on the results presented in \cite{Fedorova1, Fedorova2}. In this regard, it was necessary to conduct
a special investigation and carefully analyze foreign
observations dedicated to the search for FRBs in order
to understand the reason for their non-detection.

Table 3 was compiled, which summarizes the
observational parameters of the conducted surveys
and special observations. It shows the observational frequency (MHz), the receiver band (MHz), the
sampling interval (s), the total duration of observations (h), the area in the sky (sq. deg.) covered by the
survey, and the threshold sensitivity (Jy). It immediately draws attention to the fact that the total duration
of observations with the LPA $\sim$ 50000 h is orders of magnitude higher than all previous observations
(about 6 $\cdot$ 10$^5$ individual scans were analyzed). From
this parameter and the total number of pulses detected
with the LPA, it can be easily estimated that it takes
abou 10$^3$ hours or 10$^4$ scans on our antenna to detect a single pulse.  Another parameter that strongly distinguishes the LPA radio telescope from other instruments is the threshold sensitivity of the radio telescope
(Jy). In many works, the limiting fluence (Jy ms) is
given. It was converted into a threshold sensitivity
based on the specified sampling time or pulse duration. The sensitivity of radio telescopes of foreign colleagues is orders of magnitude worse than the sensitivity of the LPA.

In addition, if it is not possible to detect pulses with
well-tested methods, this means that non-standard
methods are necessary, such as the correlation technique used by the authors, which makes it possible to
increase the signal-to-noise several times (although
we note that the correlation technique is also the
standard method in radar). We are confident that
when the sensitivity and duration of observations of
foreign low-frequency radio telescopes reaches the LPA level, then FRBs will be detected with these
instruments as well.

The main results of this work:
\begin{enumerate}
\item The shapes of the $DM$,distribution constructed
for detected pulse signals at 111 MHz and FRBs at 1400 MHz coincide within the error and are described
by the formula for the lognormal distribution with the
parameters $\mu = 6.2$, $\sigma = 0.7$.

\item The dependence of the scattering $\tau_{sc}$ on the dispersion measure $DM$ for pulse signals recorded at a frequency of 111 MHz and FRB at 1400 MHz is constructed. The exponent $k = 0.49 \pm 0.18$ of the dependence $\tau_{sc} \sim DM^k$ within the error coincides with the
exponent $k = 0.43 \pm 0.15$ of dependence $\tau_{sc}(DM)$ for
pulses at 1.4 GHz.

\item The obtained power dependence is weaker in
comparison with the dependence for pulsars and is
explained by the fact that the matter of the intergalactic medium can be considered almost homogeneous
compared to the matter of the interstellar medium.
Thus, the intergalactic medium does not significantly
contribute to the scattering of pulses. The main contribution to the scattering of pulses is made by the matter
of the host galaxy and the galaxy in which the observer
is located.

\item The scattering value at 111 and 1400 MHz is
described by the law $\tau_{SC}(f) \sim f^{-1.9\pm 0.7}$ and, thus, differs from the dependence $f^{-4}$ derived for pulsars.

\item An estimate was made of the total component $DM_{halo} + \frac {DM_{host}}{1+z} \sim 60$\thickspace pc/cm$^{-3}$, which is determined
by the contribution of matter from the host galaxy and
the halo of our Galaxy to the dispersion measure of
radio bursts. This value depends on the model of the
distribution of matter in the Galaxy and can be corrected in the future.

\item Based on the Log $N$ -- Log $S$ dependencies constructed for detected pulses at 111 MHz and FRB at
1400 MHz, the average spectral index $\alpha = - 0.63 \pm 0.20$  is derived under the assumption of equality of the statistical properties of these samples.

\item Analysis of the form of Log $N$ -- Log $S$ dependencies
at two frequencies shows that for $F_{111 Œƒæ} > 200$ Jy ms and $F_{1.4 ƒƒæ} > 50$ Jy ms, both of them follow the law $S^{-3/2}$.

\item According to the totality of the detected observational signs: equality of $DM$ distributions, the exponent $k$ of the dependence $\tau_{sc} \sim DM^k$ coinciding
within the error, and the following the law $S^{-3/2}$ of the Log $N$ -- Log $S$ curves at frequencies of 111 and 1400 MHz, we associate the detected pulse signals
with FRBs. Thus, a frequency of 111 MHz is the lowest
frequency at which radio bursts are detected.
\end{enumerate}

{Translated by T. N. Sokolova}

\newpage
\begin{center}
{\bf Table 1.} Parameters of fast radio bursts recorded at a frequency of 111 MHz
\end{center}

\begin{tabular}{|c|c|c|c|c|c|}\hline
\rule{0mm}{2mm} FRB & Coordinates,        & $DM$, pc/cm$^3$  & Scattering  & Fluence,    & S/N \\ 
                    & $\alpha$, $\delta$ &                  & value, ms                  & Jy ms & \\
\hline
FRB121029 & 00:12:00 +42.06 & 732 & 321 & 442 & 4.3 \\ \hline
FRB141216 & 00:14:00 +41.64 & 545 & 869 & 752 & 3.6 \\ \hline
FRB131030 & 00:25:00 +39.98 & 207 & 526 & 494 & 6.9 \\ \hline
FRB180321 & 00:33:00 +42.03 & 596 & 1634 & 2326 & 5.3 \\ \hline
FRB160206 & 01:01:00 +41.63 & 1262 & 1594 & 1506 & 5.6 \\ \hline
FRB140212 & 01:31:00 +30.54 & 910 & 389 & 973 & 3.6 \\ \hline
FRB151125.1 & 01:31:00 +30.98 & 273 & 1679 & 1856 & 3.3 \\ \hline
FRB151125.2 & 01:32:00 +30.98 & 273 & 1466 & 1671 & 5.3 \\ \hline
FRB151018 & 05:21:00 +33.1 & 570 & 494 & 3500 & 5.9 \\ \hline
FRB160920 & 05:34:00 +41.75 & 1767 & 423 & 1100 & 3.3 \\ \hline
FRB170606 & 05:34:00 +41.75 & 247 & 100 & 1782 & 3.0 \\ \hline
FRB180606 & 11:43:58 +25.08 & 331 & 492 & 445 & 7.0 \\ \hline
FRB180622 & 11:46:06 +37.01 & 222 & 315 & 603 & 6.0 \\ \hline
FRB180417 & 11:47:06 +24.6 & 515 & 481 & 757 & 6.3 \\ \hline
FRB180614 & 11:48:35 +27.34 & 577 & 520 & 757 & 5.3 \\ \hline
FRB180616 & 11:48:48 +39.13 & 576 & 415 & 723 & 5.8 \\ \hline
FRB180426 & 11:49:01 +35.30 & 362 & 574 & 551 & 6.5 \\ \hline
FRB180607 & 11:49:54 +30.96 & 438 & 314 & 512 & 7.0 \\ \hline
FRB180427 & 11:52:07 +26.91 & 305 & 471 & 586 & 6.7 \\ \hline
FRB180423 & 11:53:20 +30.51 & 385 & 387 & 475 & 6.4 \\ \hline
FRB180603 & 11:56:29 +22.71 & 1865 & 1281 & 1627 & 5.9 \\ \hline
FRB180417.J1155+4112 & 11:55:27 +41.21 & 273 & 866 & 590 & 5.7 \\ \hline
FRB180627 & 11:55:58 +38.69 & 1740 & 1205 & 2456 & 7.0 \\ \hline
FRB180513 & 11:58:41 +28.27 & 750 & 947 & 620 & 5.0 \\ \hline
FRB180502 & 11:58:57 +23.66 & 570 & 533 & 690 & 5.3 \\ \hline
FRB180428 & 11:59:14 +26.50 & 375 & 358 & 534 & 9.5 \\ \hline
FRB180428.J1200+4136 & 12:00:13 +41.61 & 198 & 134 & 476 & 6.2 \\ \hline

\end{tabular}

\newpage

\begin{center}
\begin{tabular}{|c|c|c|c|c|c|} \hline
\rule{0mm}{2mm} FRB & Coordinates,        & $DM$, pc/cm$^3$  & Scattering  & Fluence,    & S/N \\ 
                    & $\alpha$, $\delta$ &                  & value, ms                  & Jy ms & \\
\hline
FRB180629 & 12:01:20 +26.50 & 307 & 280 & 352 & 6.5 \\ \hline
FRB180507 & 12:03:18 +41.62 & 625 & 792 & 985 & 6.9 \\ \hline
FRB180429 & 12:03:32 +40.79 & 348 & 503 & 551 & 7.3 \\ \hline
FRB180502.J1207+3726 & 12:07:07 +40.79 & 1373 & 2612 & 2292 & 6.5 \\ \hline
FRB180625 & 12:07:23 +33.59 & 245 & 190 & 273 & 6.3 \\ \hline
FRB180628 & 12:08:20 +22.71 & 300 & 302 & 414 & 6.4 \\ \hline
FRB180609 & 12:08:56 +29.19 & 324 & 269 & 357 & 6.9 \\ \hline
FRB180616.J1210+2722 & 12:10:50 +27.37 & 560 & 642 & 654 & 6.6\\ \hline
FRB180617 & 12:11:25 +34.02 & 575 & 453 & 586 & 6.2 \\ \hline
FRB180521 & 12:12:10 +27.82 & 214 & 255 & 487 & 6.1 \\ \hline
FRB180507.J1212+2116 & 12:12:19 +21.28 & 560 & 256 & 916 & 6.6 \\ \hline
FRB180502.J1216+3750 & 12:16:51 +37.85 & 638 & 528 & 729 & 8.5 \\ \hline
FRB180503 & 12:18:42 +27.30 & 242 & 561 & 633 & 7.0 \\ \hline
FRB180531.J1221+3751 & 12:21:26 +37.85 & 465 & 519 & 679 & 4.9 \\ \hline
FRB180603.J1223+3726 & 12:23:11 +37.44 & 1680 & 1165 & 1037 & 4.3 \\ \hline
FRB180504 & 12:25:51 +41.21 & 670 & 626 & 502 & 5.0 \\ \hline
FRB180514 & 12:26:59 +34.44 & 288 & 305 & 348 & 5.6 \\ \hline
FRB180604 & 12:27:08 +32.28 & 219 & 632 & 403 & 5.9 \\ \hline
FRB180522 & 12:27:43 +26.91 & 578 & 217 & 432 & 6.2 \\ \hline
FRB180504.J1228+2844 & 12:28:19 +28.74 & 439 & 764 & 463 & 6.3 \\ \hline
FRB180521.J1228+4112 & 12:28:49 +41.21 & 279 & 220 & 290 & 7.0 \\ \hline
FRB180516 & 12:29:19 +38.70 & 170 & 184 & 389 & 6.0 \\ \hline
FRB180509.J1229+3030 & 12:29:37 +30.50 & 231 & 744 & 491 & 5.8 \\ \hline
FRB180605 & 12:29:47 +29.19 & 227 & 560 & 536 & 11.5 \\ \hline
FRB180609.J1230+2627 & 12:30:19 +26.46 & 420 & 913 & 1018 & 8.0 \\ \hline
FRB180607.J1231+2911 & 12:31:20 +29.19 & 350 & 431 & 583 & 6.8 \\ \hline
FRB180610 & 12:31:46 +4121 & 175 & 508 & 372 & 5.7 \\ \hline
\end{tabular}
\end{center}

\newpage

\begin{center}
\begin{tabular}{|c|c|c|c|c|c|} \hline
\rule{0mm}{2mm} FRB & Coordinates,        & $DM$, pc/cm$^3$  & Scattering  & Fluence,    & S/N \\ 
                    & $\alpha$, $\delta$ &                  & value, ms                  & Jy ms & \\
\hline
FRB180531 & 12:31:56 +38.28 & 310 & 786 & 723 & 7.5	 \\ \hline
FRB180615 & 12:37:05 +38.70 & 450 & 597 & 633 & 6.2 \\ \hline
FRB180504.J1243+3635 & 12:43:05 +36.59 & 608 & 793 & 872 & 4.3 \\ \hline
FRB180511 & 12:43:07 +26.46 & 1049 & 1234 & 1282 & 7.0 \\ \hline
FRB180620 & 12:43:42 +41.62 & 155 & 121 & 349 & 7.0 \\ \hline
FRB180504.J1244+3518 & 12:44:11 +35.30 & 298 & 293 & 329 & 6.0 \\ \hline
FRB180620.J1245+3124 & 12:45:49 +31.40 & 409 & 244 & 459& 5.1\\ \hline
FRB180601 & 12:48:59 +24.13 & 403 & 311 & 704 & 6.5 \\ \hline
FRB161202 & 23:44:00 +40.80 & 291 & 808 & 705 & 4.2 \\ \hline
\end{tabular}
\end{center}

\newpage
\begin{center}
{\bf Table 2.} Break frequency $f_1$ at different spectral indices $\alpha$.
\end{center}

\begin{center}
\begin{tabular}{|c|c|}
\hline
\rule{0mm}{5mm} $\alpha$ & $f_1, MHz$ \\
\hline
\rule{0mm}{5mm}
-1  & 130 \\
-1.3  & 230 \\
-1.7  & 350 \\
\hline
\end{tabular}
\end{center}

\newpage
\begin{center}

{\bf Table 3.} Observational parameters of the conducted surveys and special observations of fast radio bursts.
\end{center}

\begin{tabular}{|c|c|c|c|c|c|c|c|c|}
\hline
\rule{0mm}{1mm} Telescope & Frequency       & Band, & Sampling & $'_{obs.}$, h & Area & $S_{fl}$, & Article & Note \\
                         & of obs.,   & MHz     & interval  & of search,        & sq. deg.       &  Jy             &            & \\
                         & MHz           &         & $\tau$, s&                &          &       &             &             \\
\hline
\rule{0mm}{1mm}
MWA  & 170 - 200 & 1.28x24 & 0.5 & ? & 450 & 0.84, & Sokolowski,  & Sensitivity   \\ 
     &           &         &     &   &     & 4.57, & 2018 \cite{Sokolowski} & for different  \\         
     &           &         &     &   &     & 6.64  &  & sources\\
\hline
MWA & 139 - 170 & 1.28x24 & 2 & 10.5 & 400 & 0.35 & Tingay,  & \\ 
 &           &         &     &   &     &   & 2015 \cite{Tingay} & \\
\hline
LOFAR  & 145 & 6 & 0.005 & 1445 & 4193 & 62 & Karastergiou,  & \\
(UK)&           &         &     &   &     &   & 2015 \cite{Karastergiou} & \\
 \hline
LOFAR & 110 - 190 & 80 & 0.05 & 2x0.67 & - & 2 & Chawla,  & FRB180916. \\ 
&           &         &     &   &     &   & 2020 \cite{Chawla} & J0158+65\\
\hline
LOFAR & 110 - 188 & 78 & 0.004 & 18.3 & 0.007 & 10 & Houben,  & FRB121102 \\ 
&           &         &     &   &     &   & 2019 \cite{Houben} & \\
\hline
'€ & 111 & 2.5 & 1 & 49910 & 310 & 0.044 & Fedorova, & 0.14 Jy$/\sqrt{10}$ \\
&           &         &     &   &     &   & Rodin, & \\
&           &         &     &   &     &   & 2019 \cite{Fedorova1, Fedorova2} & \\
\hline
\end{tabular}

\newpage

\newpage

\newpage

\newpage

\newpage

\newpage

\newpage
\begin{figure}[h]
	\setcaptionmargin{1mm}
	\vbox{\includegraphics[width=1\linewidth]{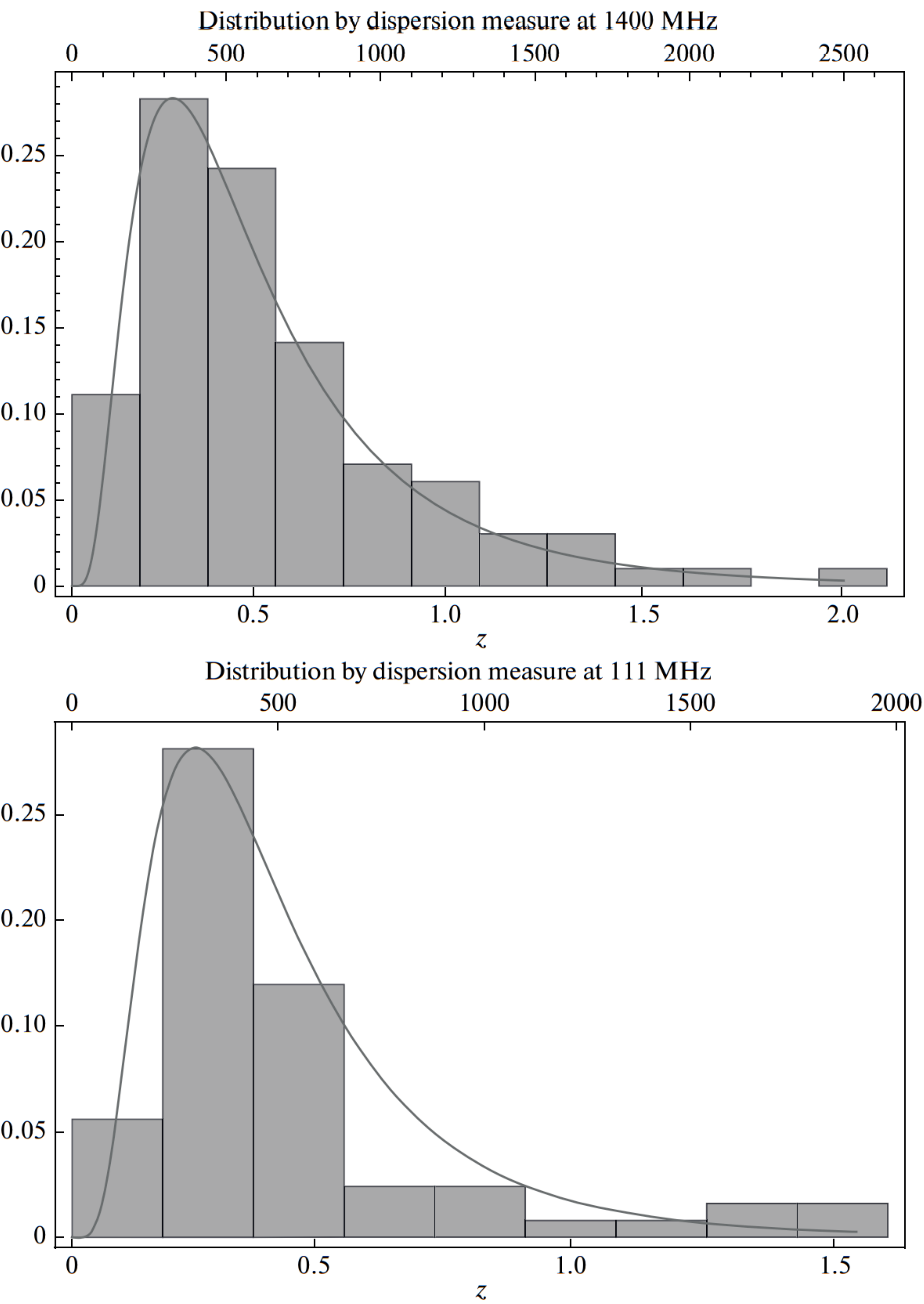}}
	\caption{Histogram of the $DM$ distribution of fast radio bursts.}
	\label{ris:fig7}
\end{figure}

\newpage
\begin{figure}[h!]
	\setcaptionmargin{1.3mm}
	\vbox{\includegraphics[width=0.95\linewidth]{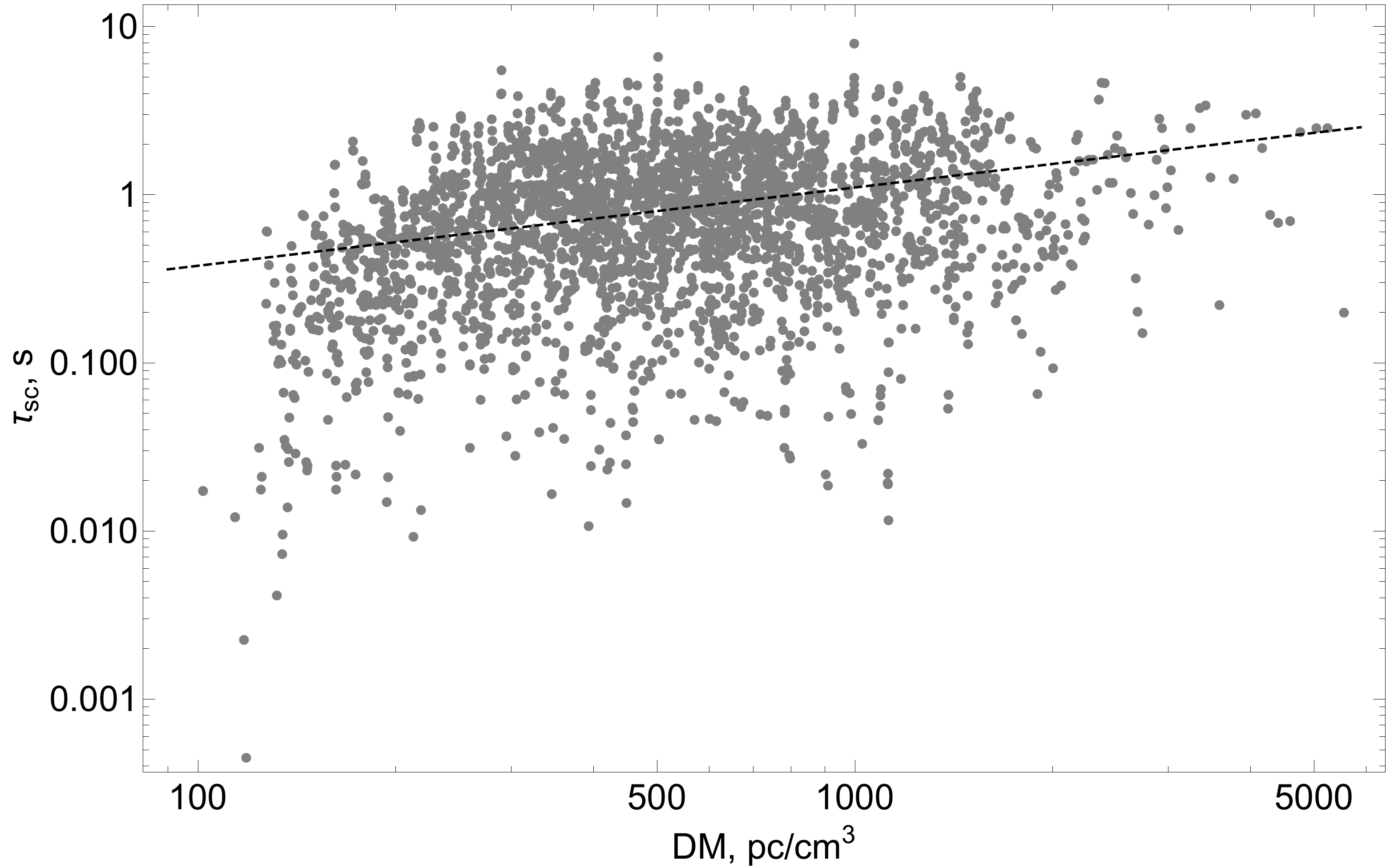}}
	\caption{The result of modeling the scattering of pulses by the dispersion measure. The position of the screen along the line of sight
	relative to the observer and the area of occurrence of a fast radio burst were selected so that the exponent $k$ coincided with the
	experimental ones. The value $\tau_{sc}$ is given in seconds.}
	\label{ris:fig8}
\end{figure}

\newpage
\begin{figure}[h!]
	\setcaptionmargin{1.3mm}
	\vbox{\includegraphics[width=1\linewidth]{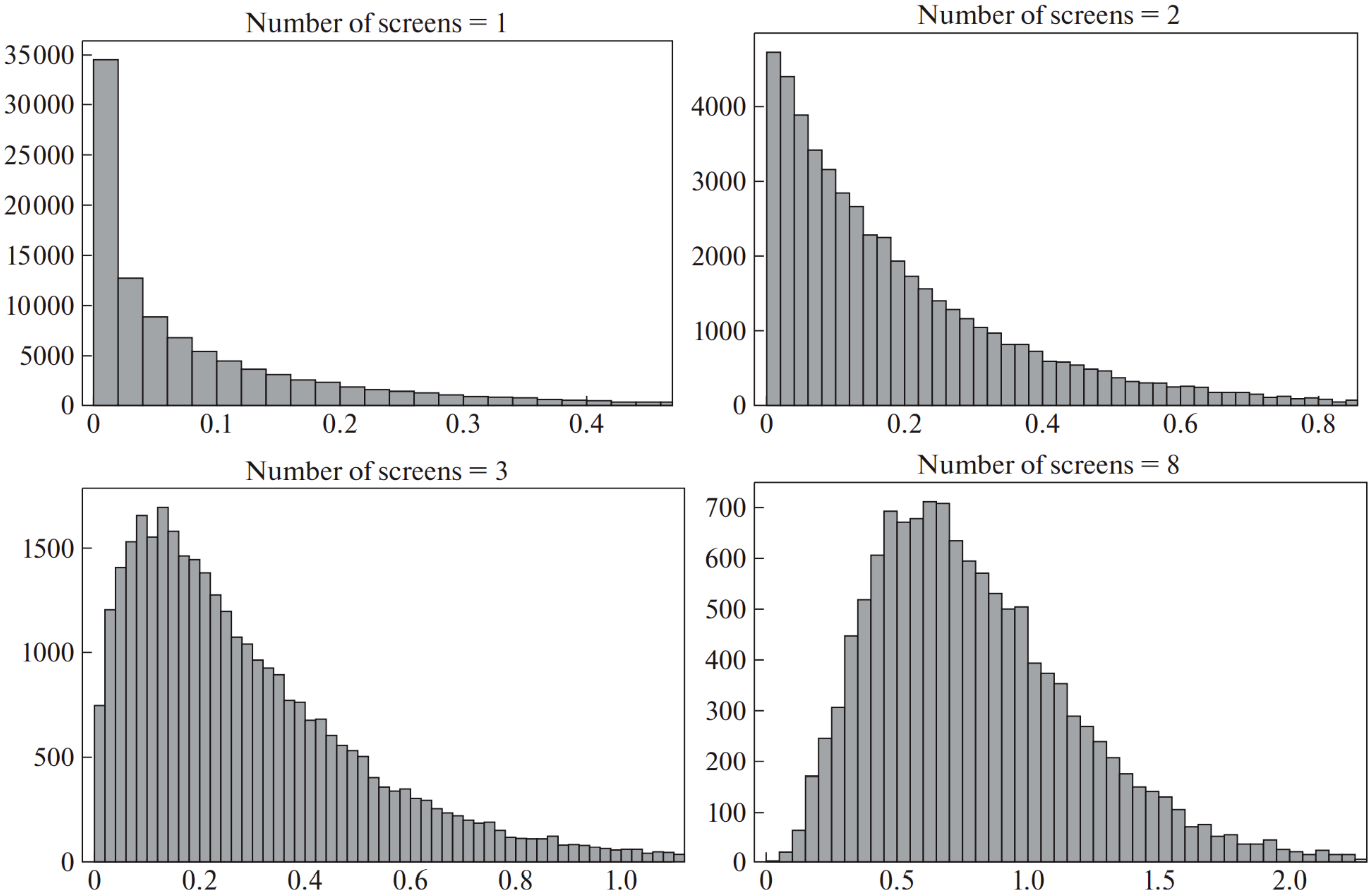}}
	\caption{Dependence of the pulse shape of a fast radio burst on the number of scattering screens on the line of sight.}
	\label{ris:fig9}
\end{figure}

\newpage
\begin{figure}[h!]
	\setcaptionmargin{1.3mm}
	\vbox{\includegraphics[width=1\linewidth]{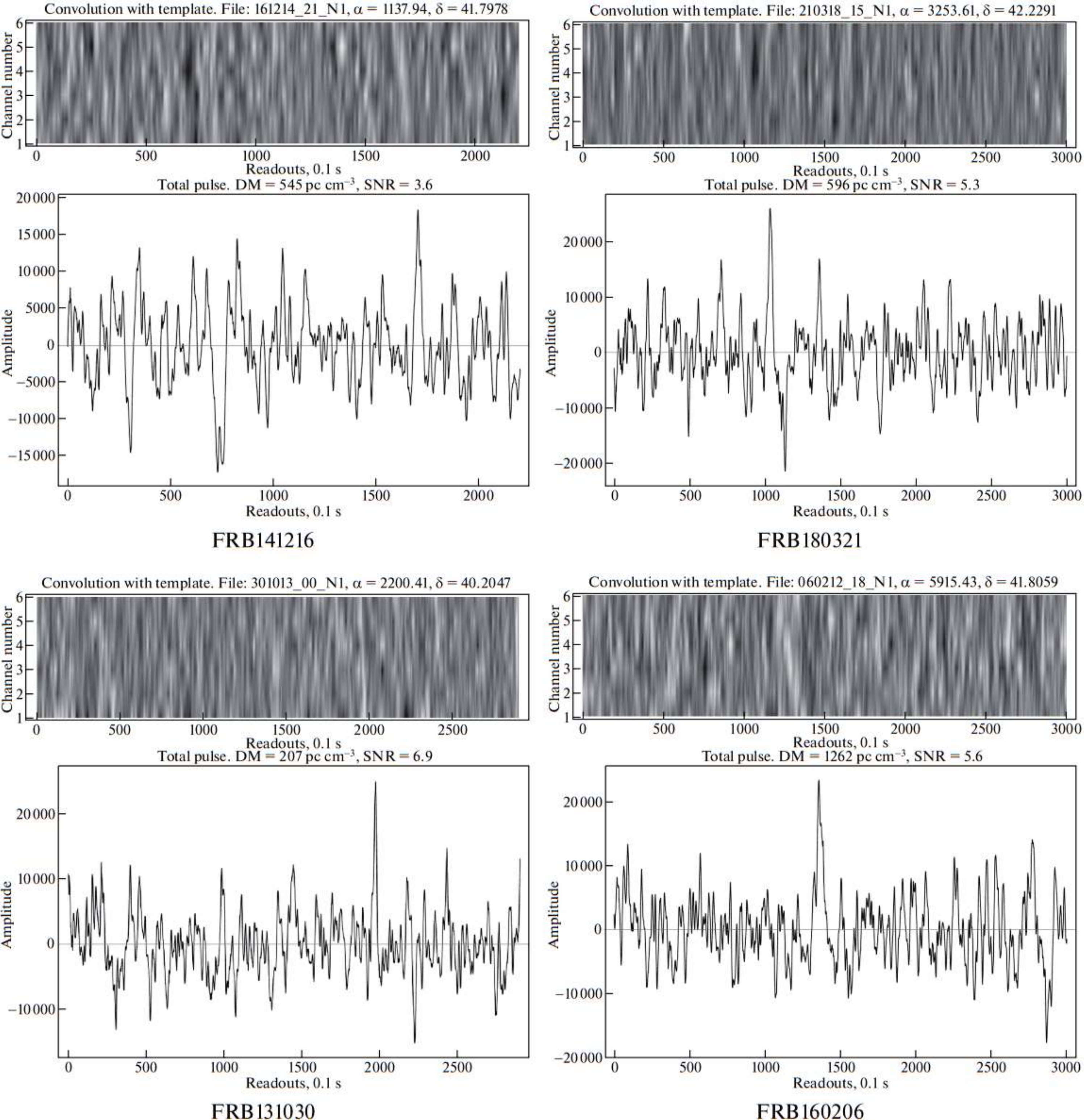}}
	\caption{Dynamic spectra and profiles of pulses.}
	\label{ris:fig10}
\end{figure}

\newpage
\begin{figure}[h!]
	\setcaptionmargin{1.3mm}
	\vbox{\includegraphics[width=1\linewidth]{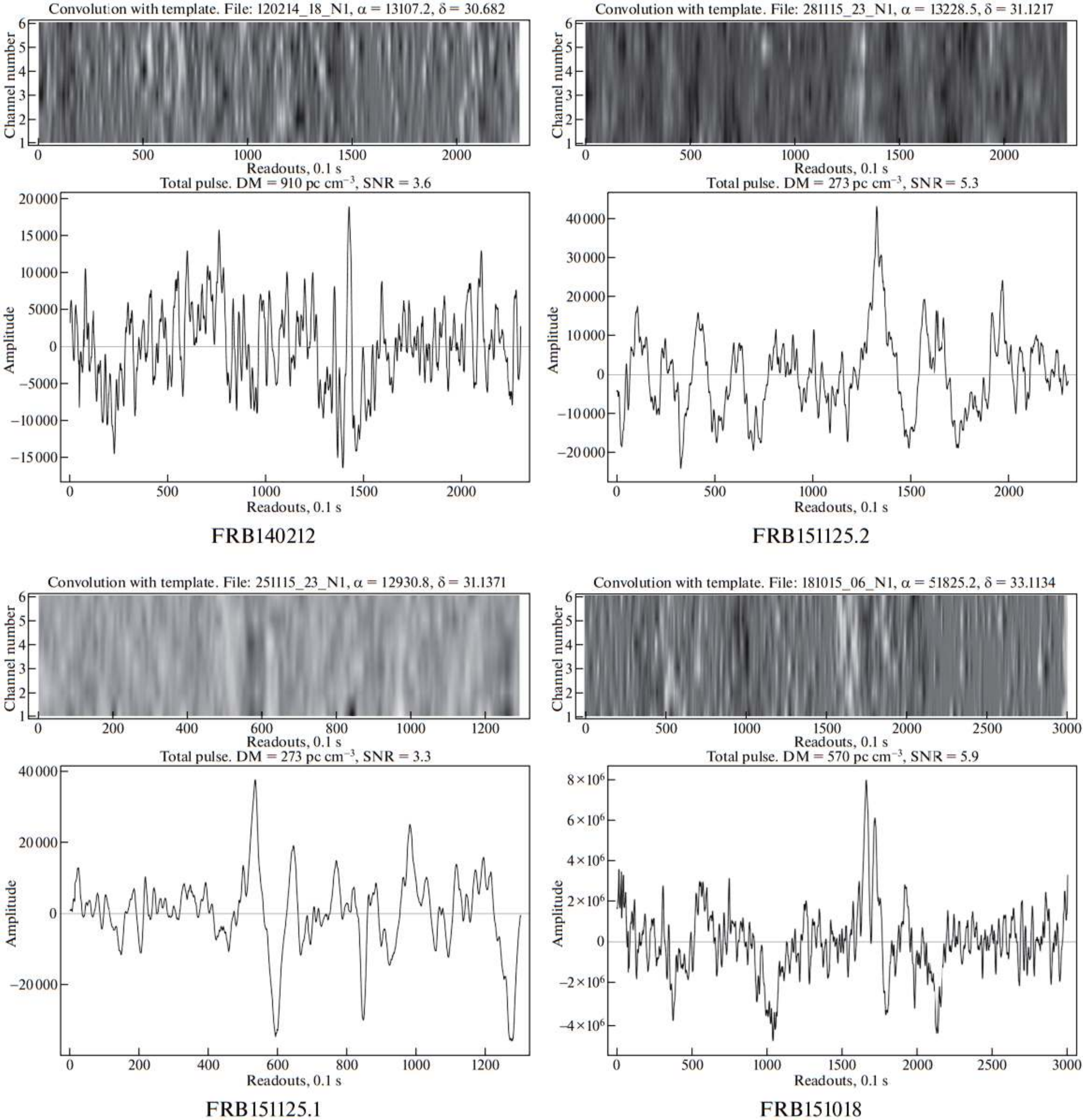}}
	\caption{Dynamic spectra and profiles of pulses.}
	\label{ris:fig11}
\end{figure}

\newpage
\begin{figure}[h!]
	\setcaptionmargin{1.3mm}
	\vbox{\includegraphics[width=1\linewidth]{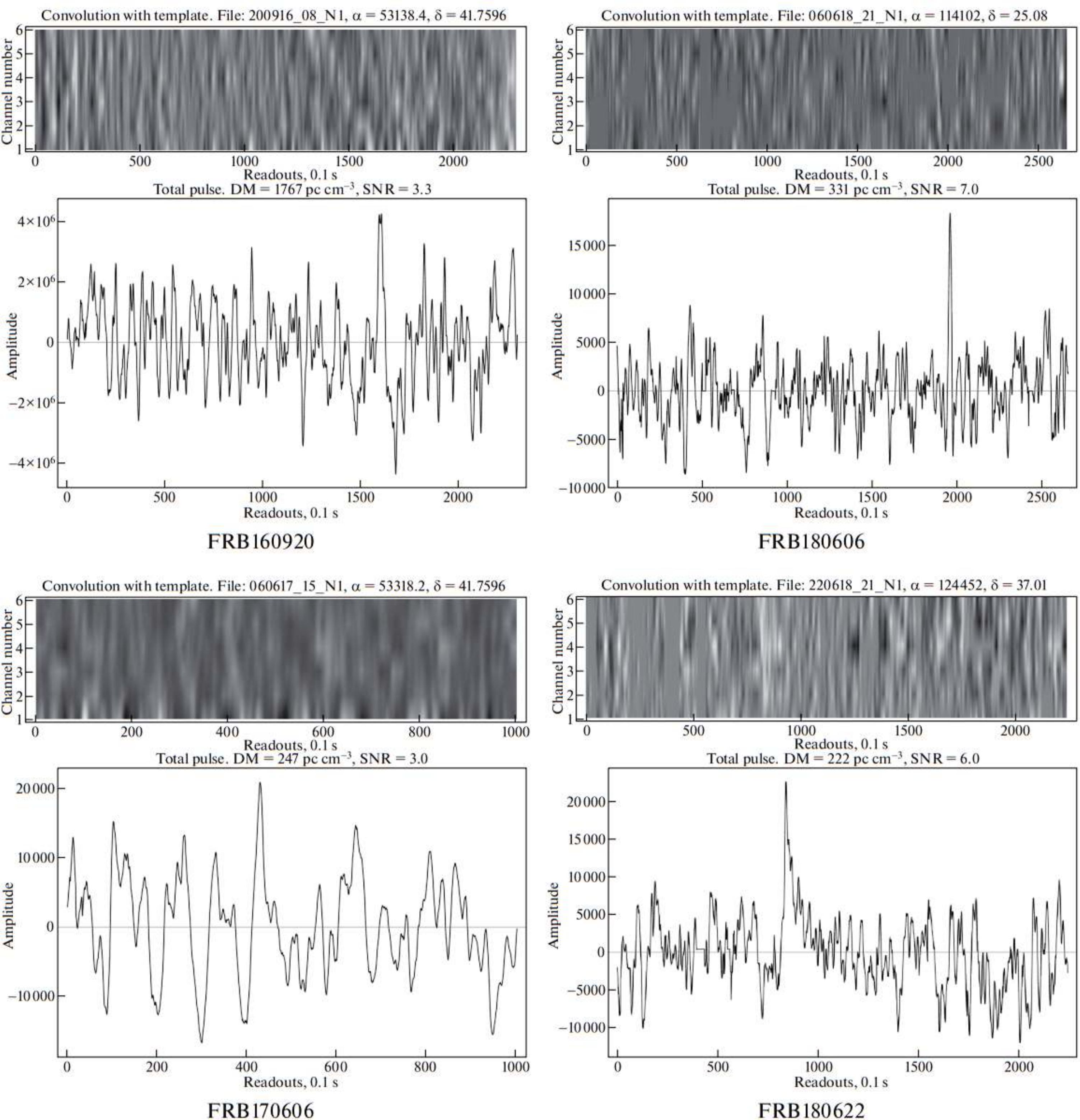}}
	\caption{Dynamic spectra and profiles of pulses.}
	\label{ris:fig12}
\end{figure}

\newpage
\begin{figure}[h!]
	\setcaptionmargin{1.3mm}
	\vbox{\includegraphics[width=1\linewidth]{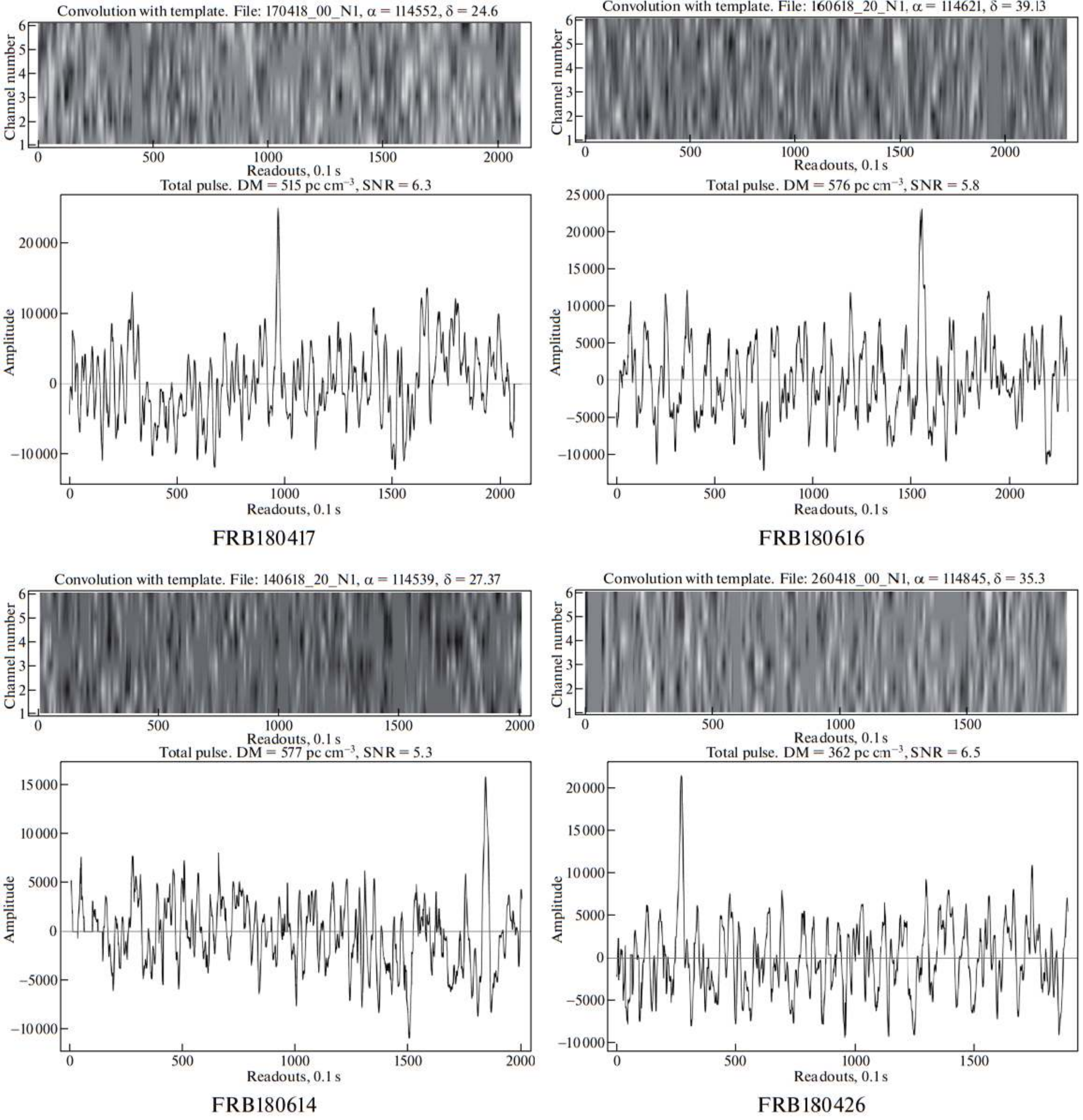}}
	\caption{Dynamic spectra and profiles of pulses.}
	\label{ris:fig13}
\end{figure}

\newpage
\begin{figure}[h!]
	\setcaptionmargin{1.3mm}
	\vbox{\includegraphics[width=1\linewidth]{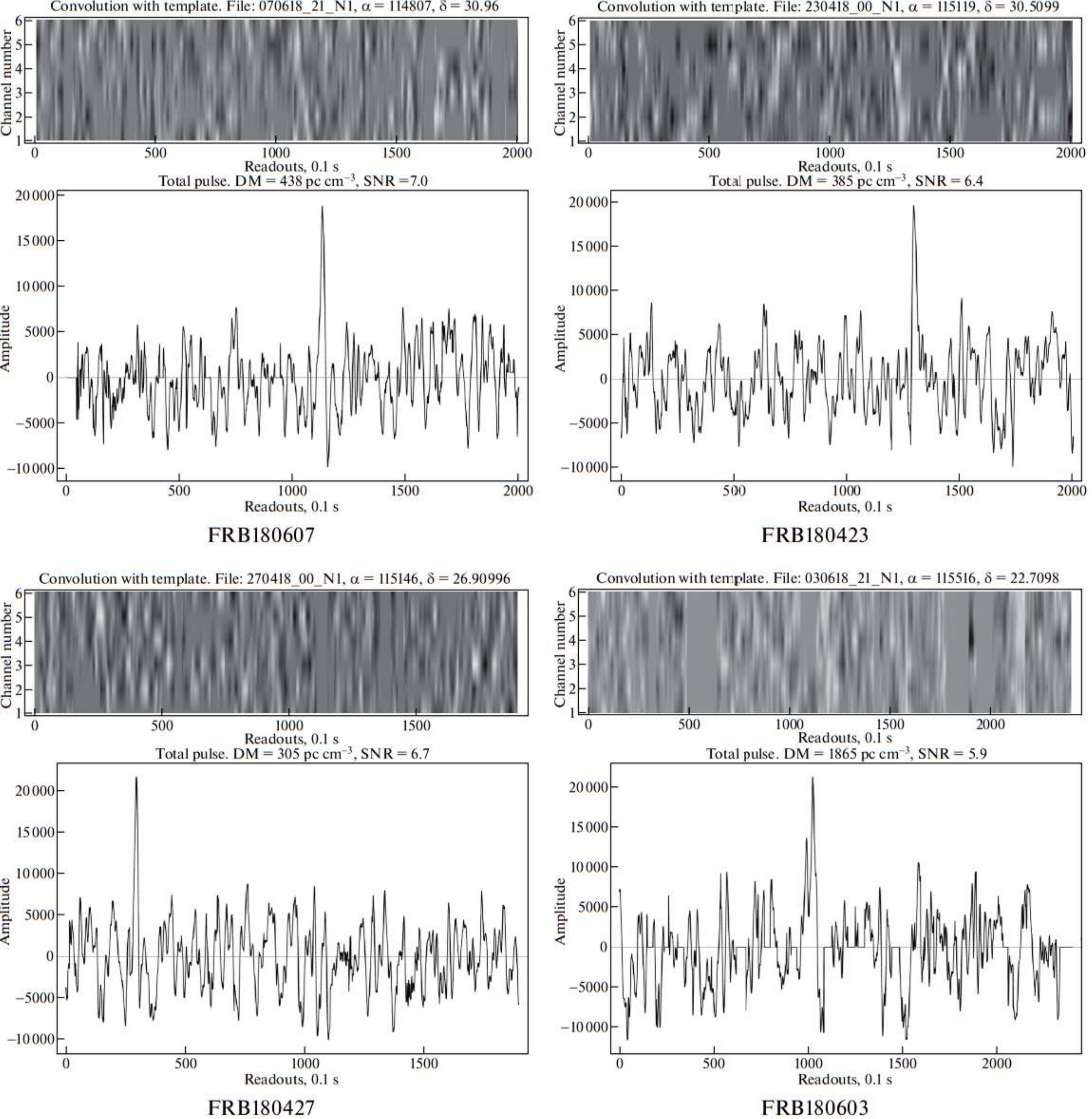}}
	\caption{Dynamic spectra and profiles of pulses.}
	\label{ris:fig14}
\end{figure}

\newpage
\begin{figure}[h!]
	\setcaptionmargin{1.3mm}
	\vbox{\includegraphics[width=1\linewidth]{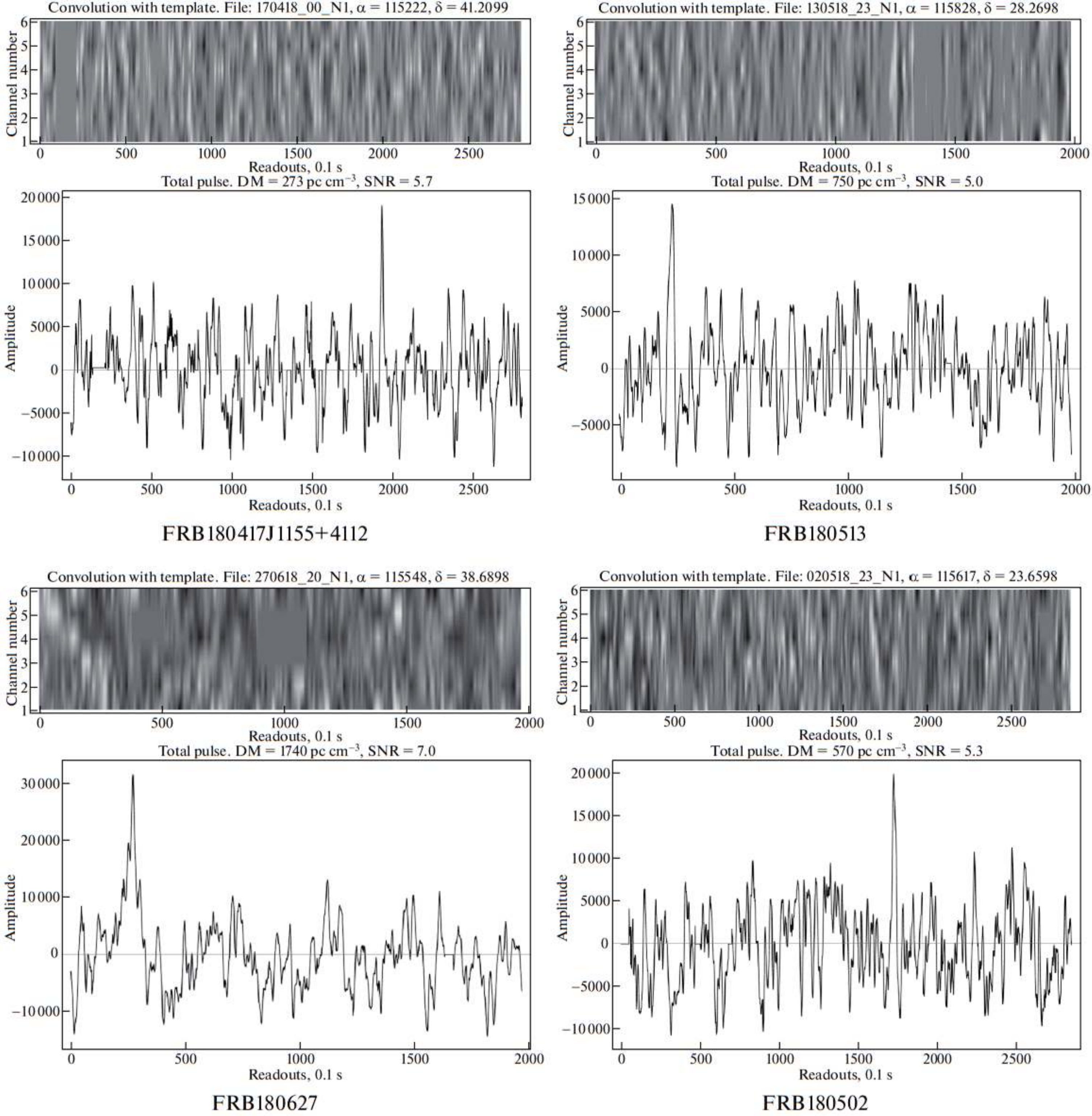}}
	\caption{Dynamic spectra and profiles of pulses.}
	\label{ris:fig15}
\end{figure}

\newpage
\begin{figure}[h!]
	\setcaptionmargin{1.3mm}
	\vbox{\includegraphics[width=1\linewidth]{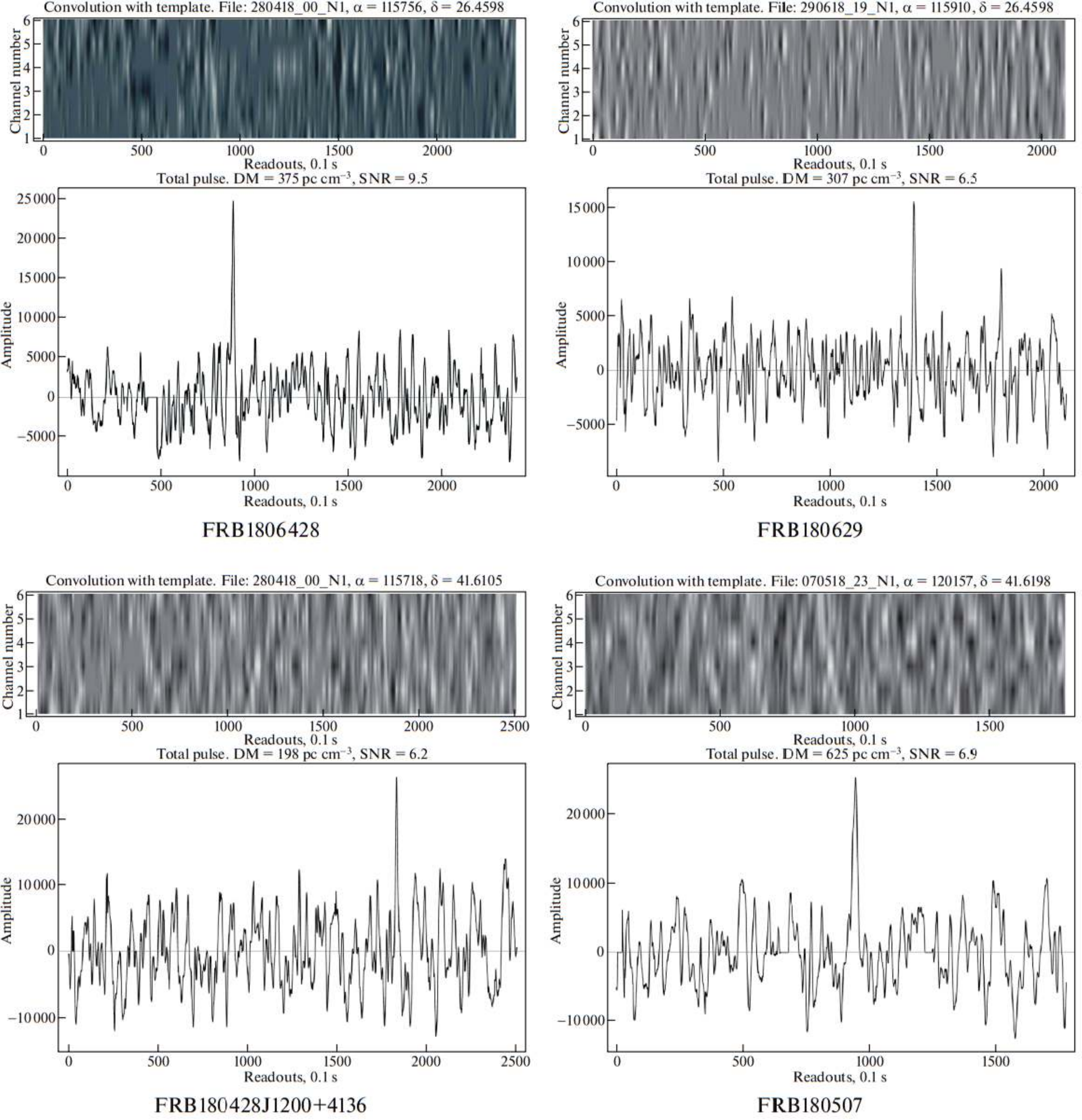}}
	\caption{Dynamic spectra and profiles of pulses.}
	\label{ris:fig16}
\end{figure}

\newpage
\begin{figure}[h!]
	\setcaptionmargin{1.3mm}
	\vbox{\includegraphics[width=1\linewidth]{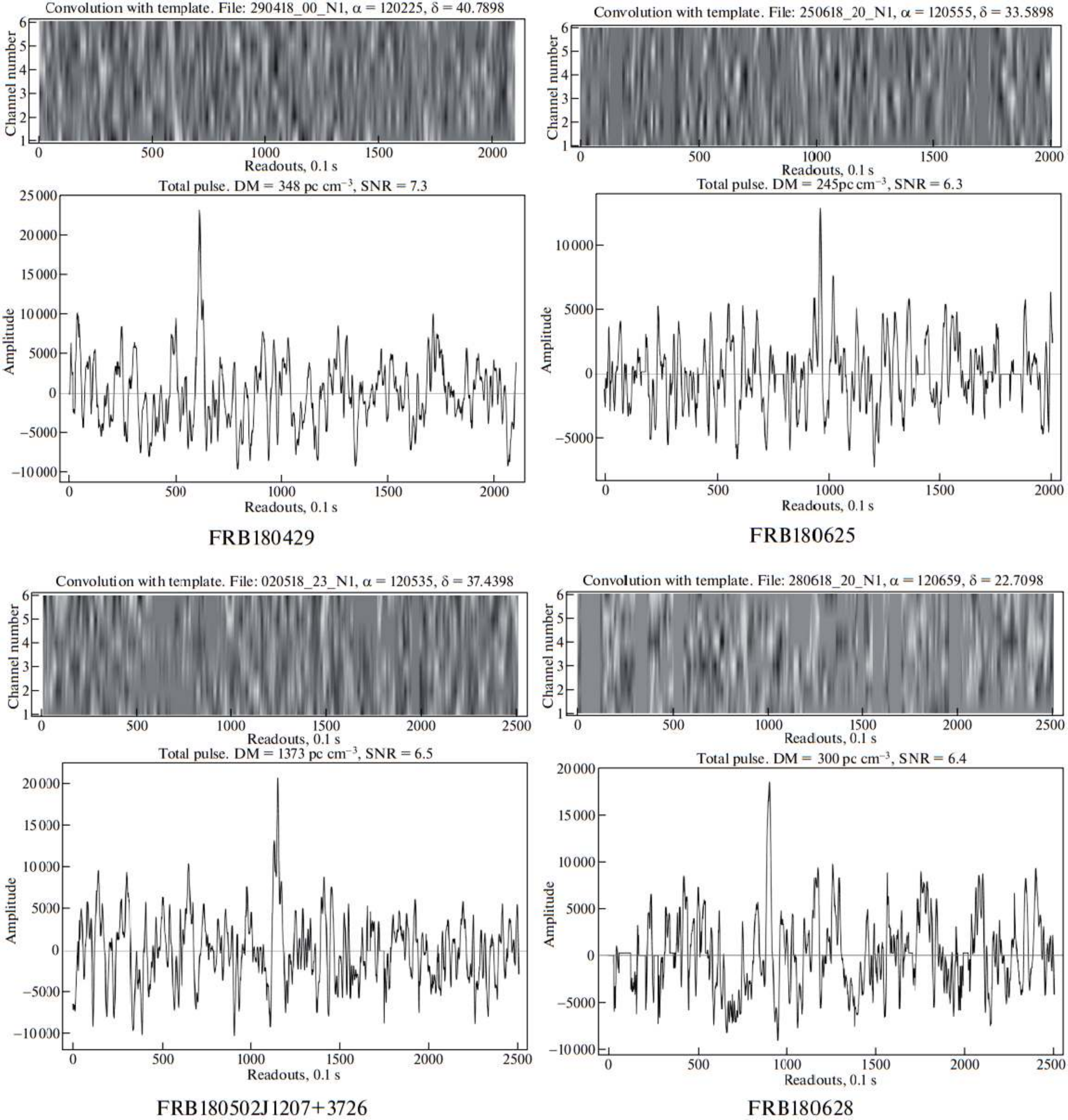}}
	\caption{Dynamic spectra and profiles of pulses.}
	\label{ris:fig17}
\end{figure}

\newpage
\begin{figure}[h!]
	\setcaptionmargin{1.3mm}
	\vbox{\includegraphics[width=1\linewidth]{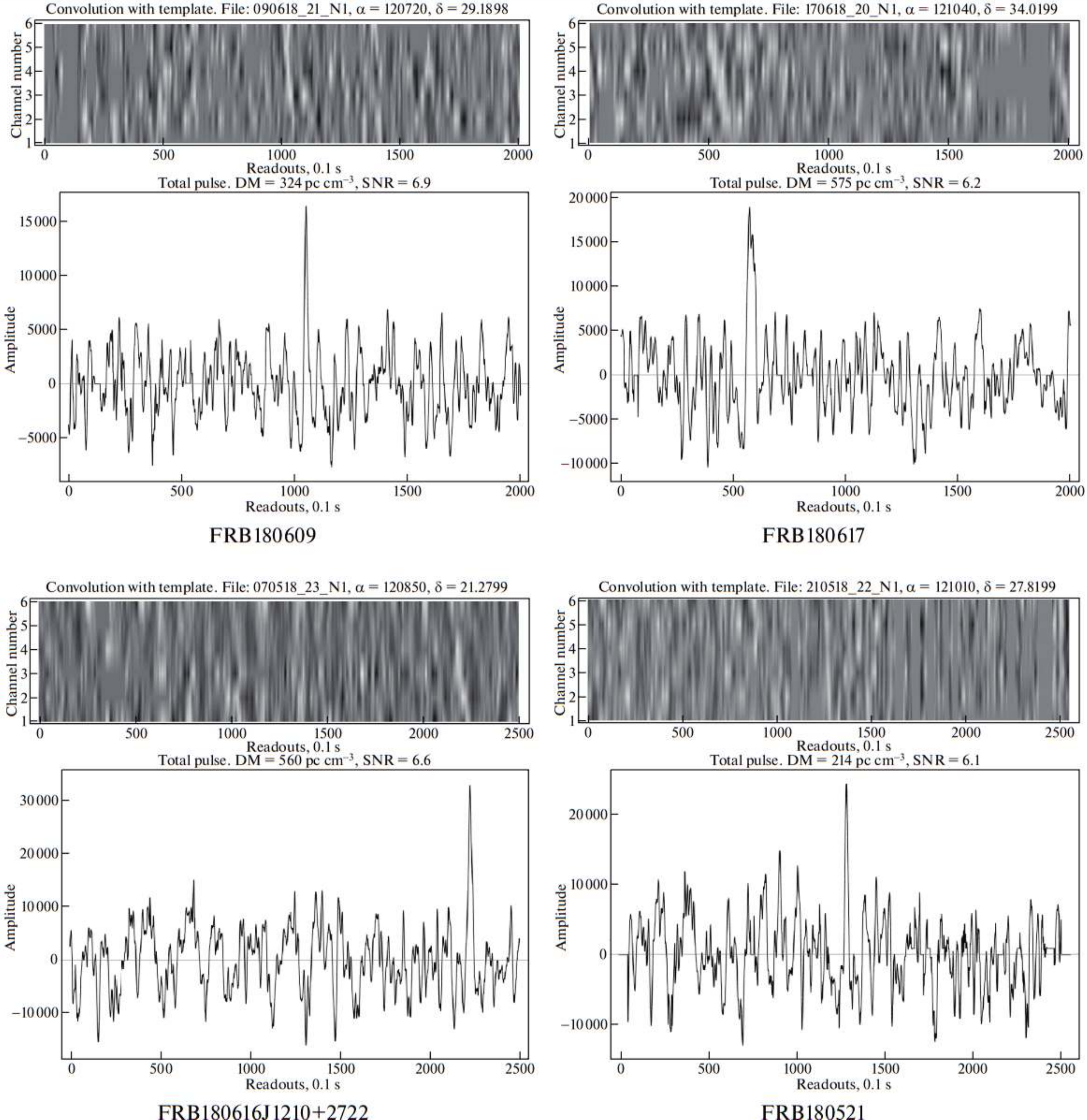}}
	\caption{Dynamic spectra and profiles of pulses.}
	\label{ris:fig18}
\end{figure}

\newpage
\begin{figure}[h!]
	\setcaptionmargin{1.3mm}
	\vbox{\includegraphics[width=1\linewidth]{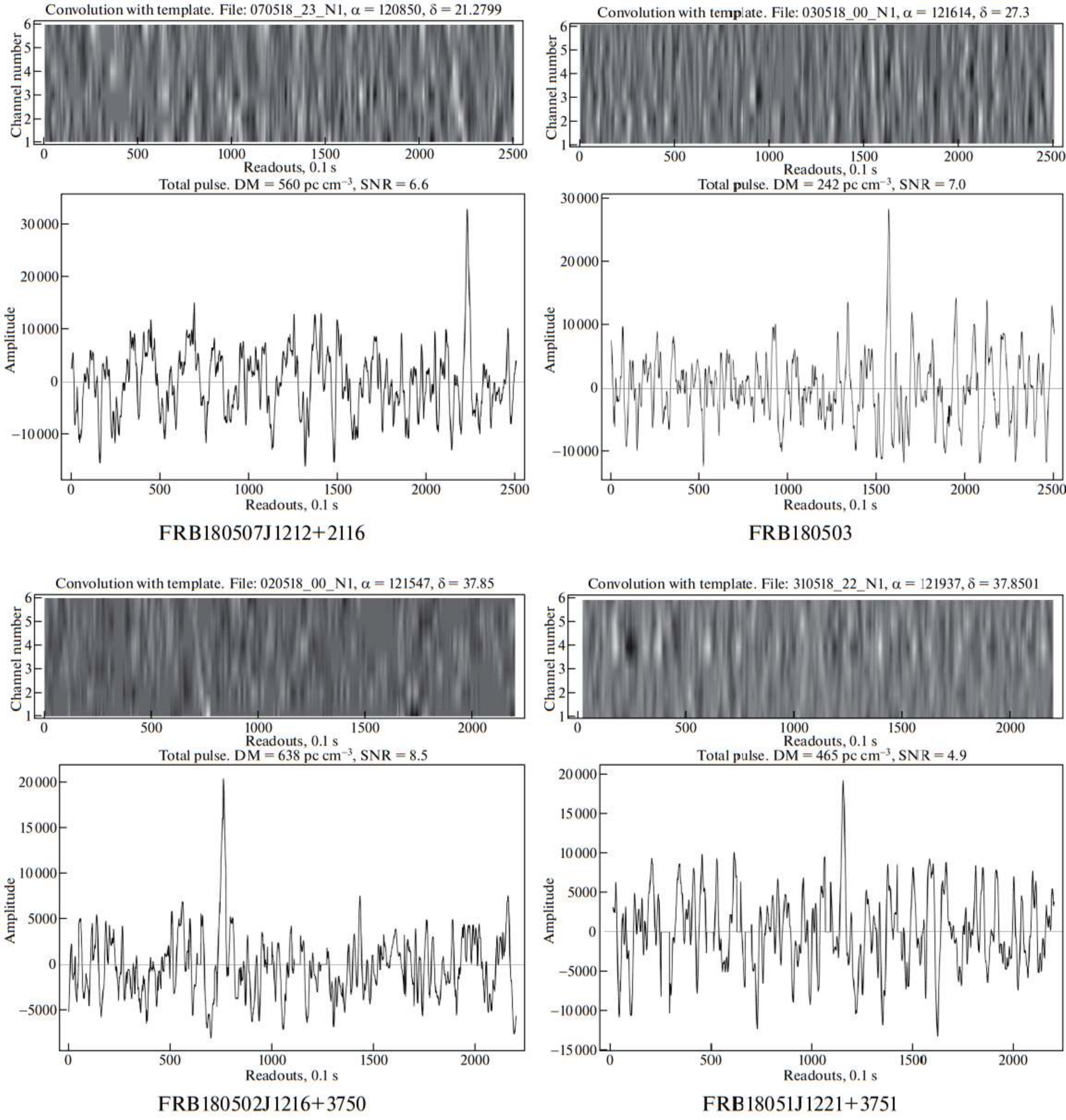}}
	\caption{Dynamic spectra and profiles of pulses.}
	\label{ris:fig19}
\end{figure}

\newpage
\begin{figure}[h!]
	\setcaptionmargin{1.3mm}
	\vbox{\includegraphics[width=1\linewidth]{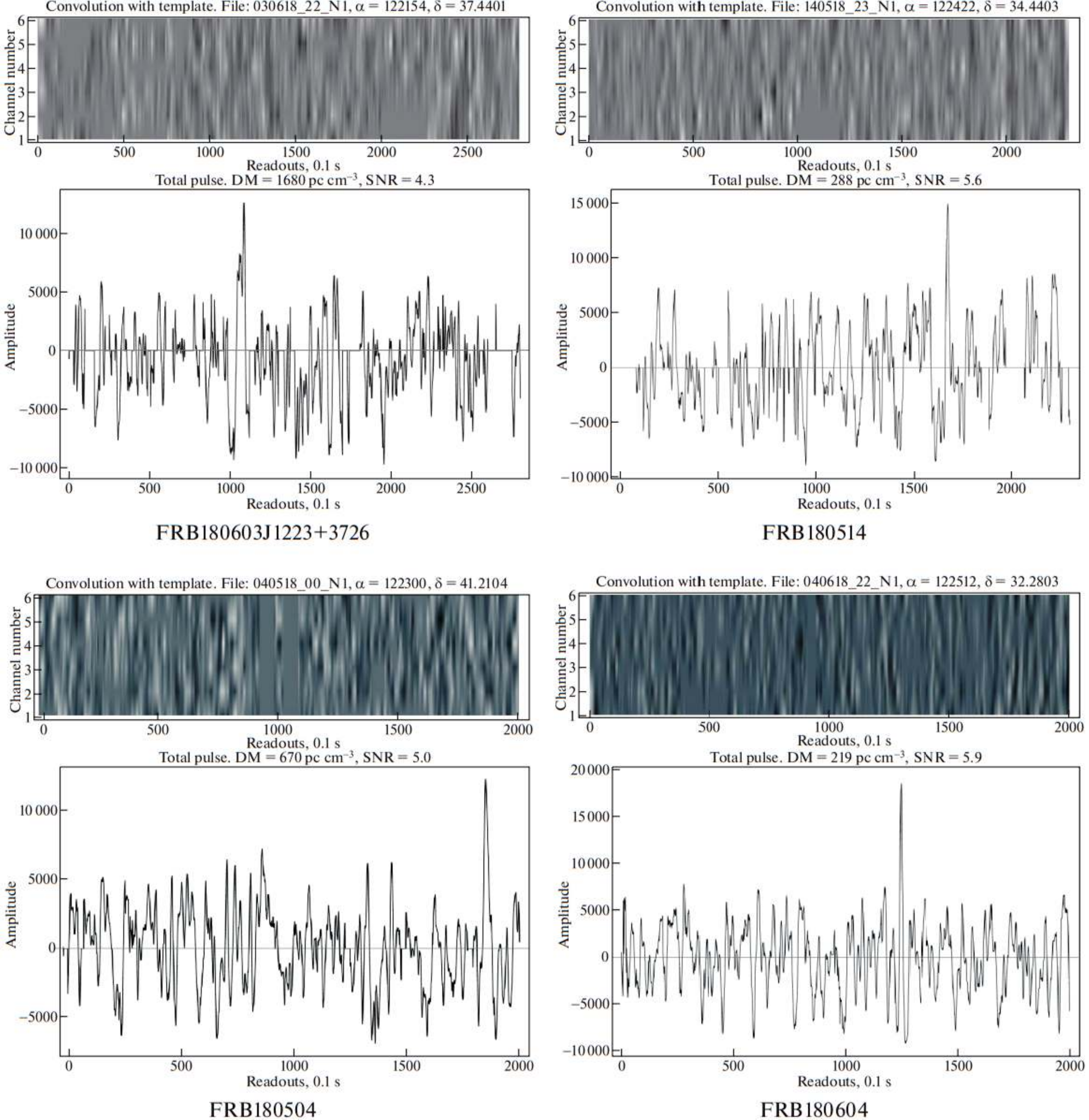}}
	\caption{Dynamic spectra and profiles of pulses.}
	\label{ris:fig20}
\end{figure}

\newpage
\begin{figure}[h!]
	\setcaptionmargin{1.3mm}
	\vbox{\includegraphics[width=1\linewidth]{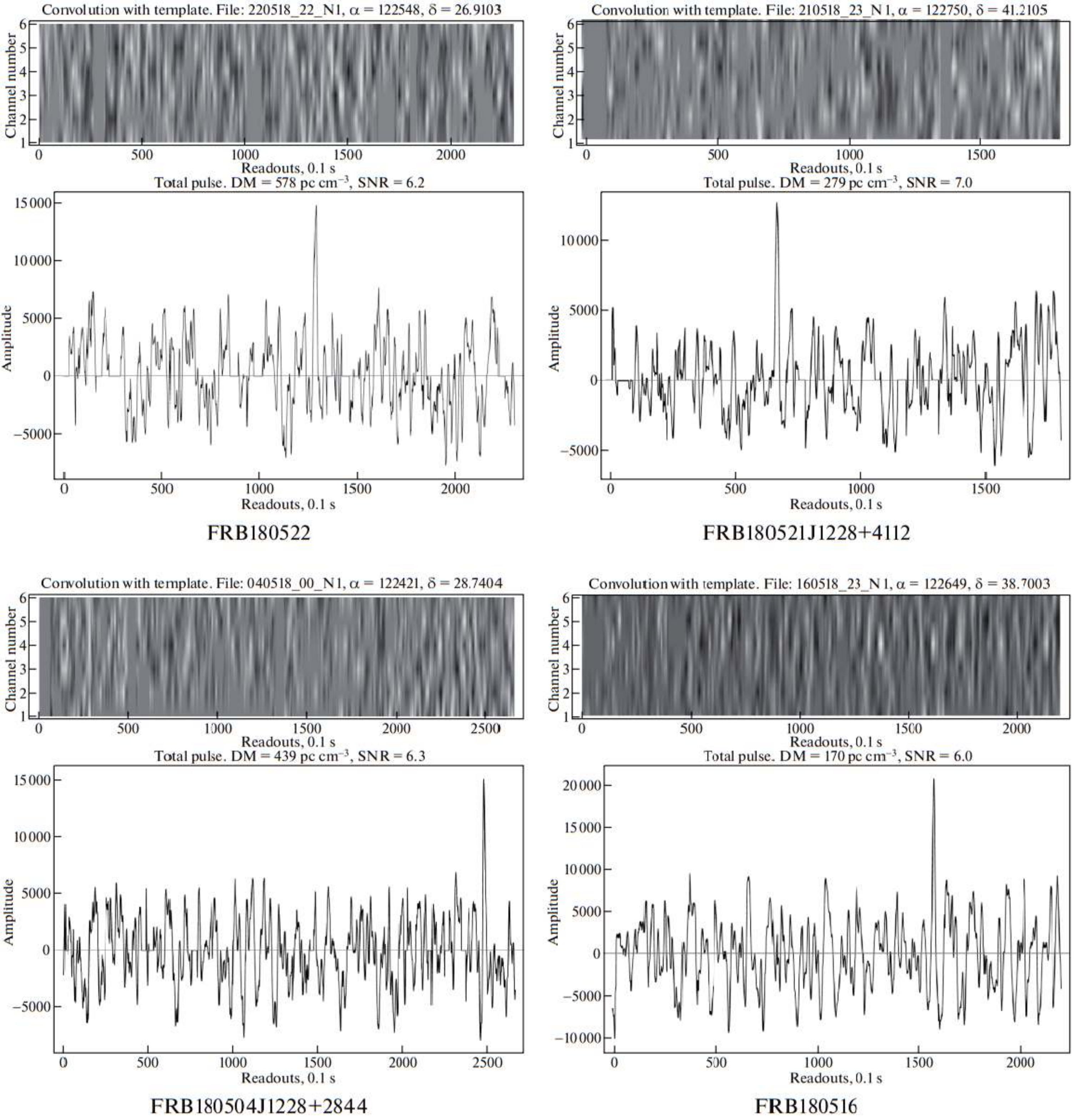}}
	\caption{Dynamic spectra and profiles of pulses.}
	\label{ris:fig21}
\end{figure}

\newpage
\begin{figure}[h!]
	\setcaptionmargin{1.3mm}
	\vbox{\includegraphics[width=1\linewidth]{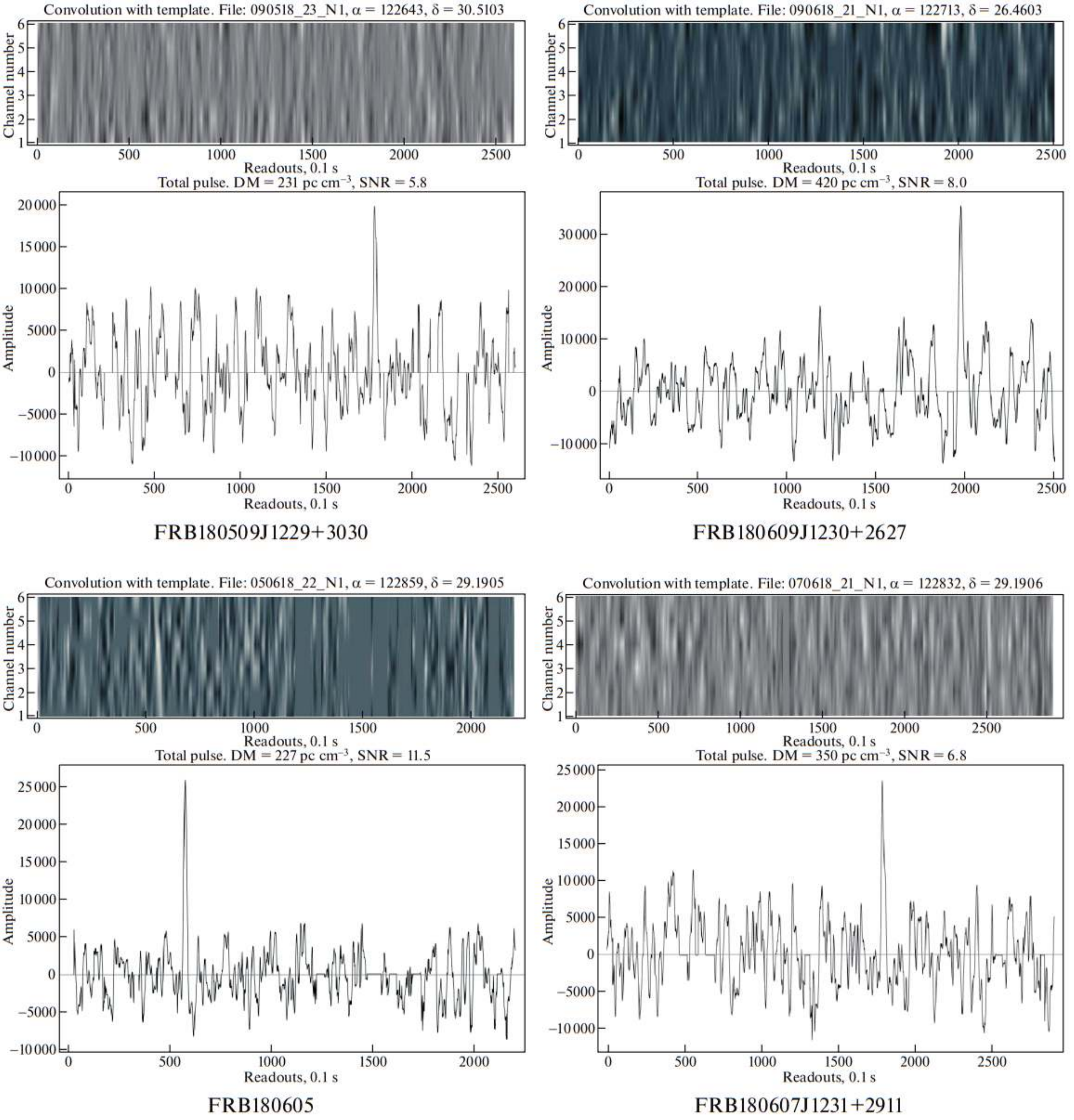}}
	\caption{Dynamic spectra and profiles of pulses.}
	\label{ris:fig22}
\end{figure}

\clearpage
\begin{figure}[h!]
	\setcaptionmargin{1.3mm}
	\vbox{\includegraphics[width=1\linewidth]{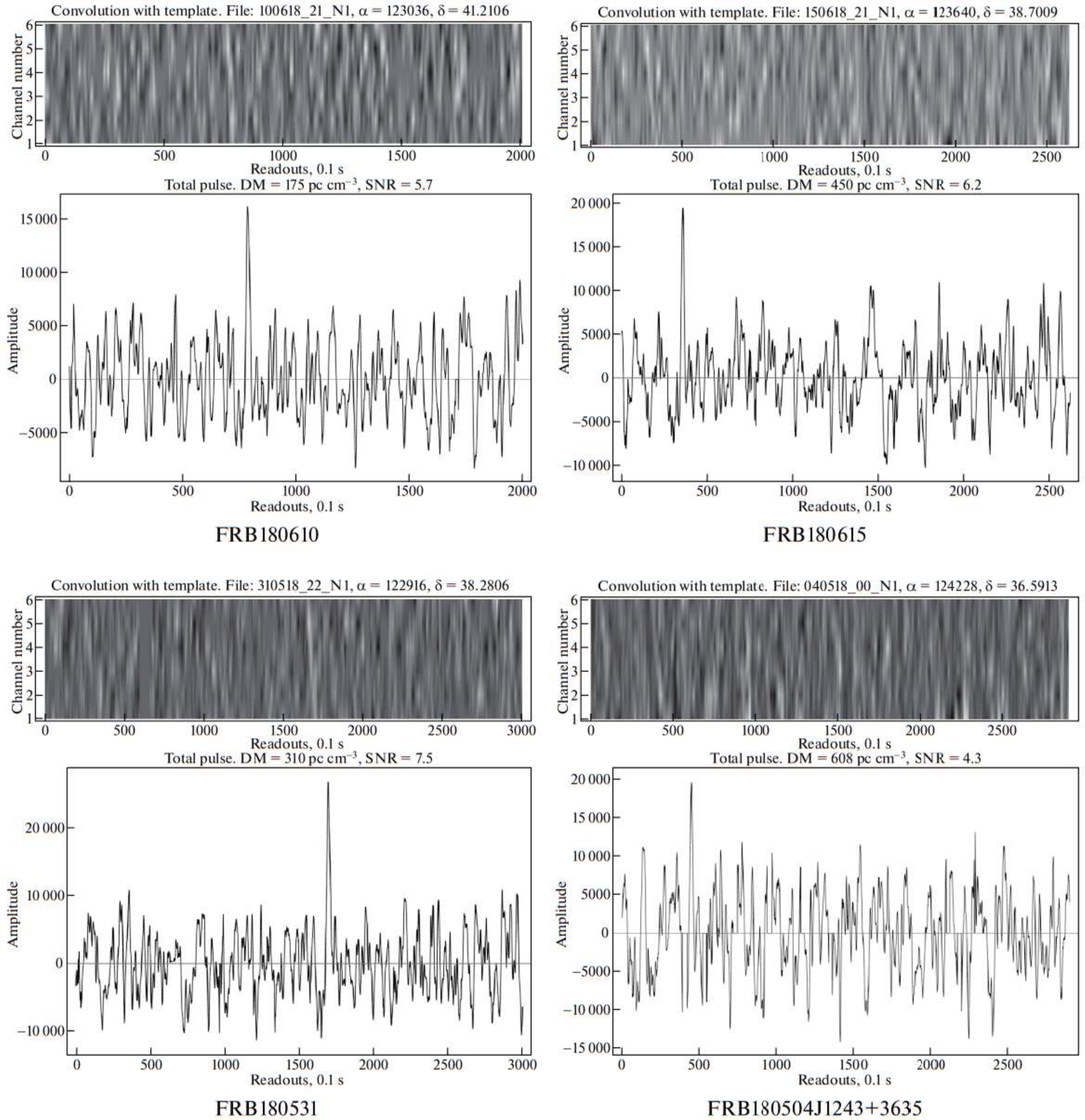}}
	\caption{Dynamic spectra and profiles of pulses.}
	\label{ris:fig23}
\end{figure}

\newpage
\begin{figure}[h!]
	\setcaptionmargin{1.3mm}
	\vbox{\includegraphics[width=1\linewidth]{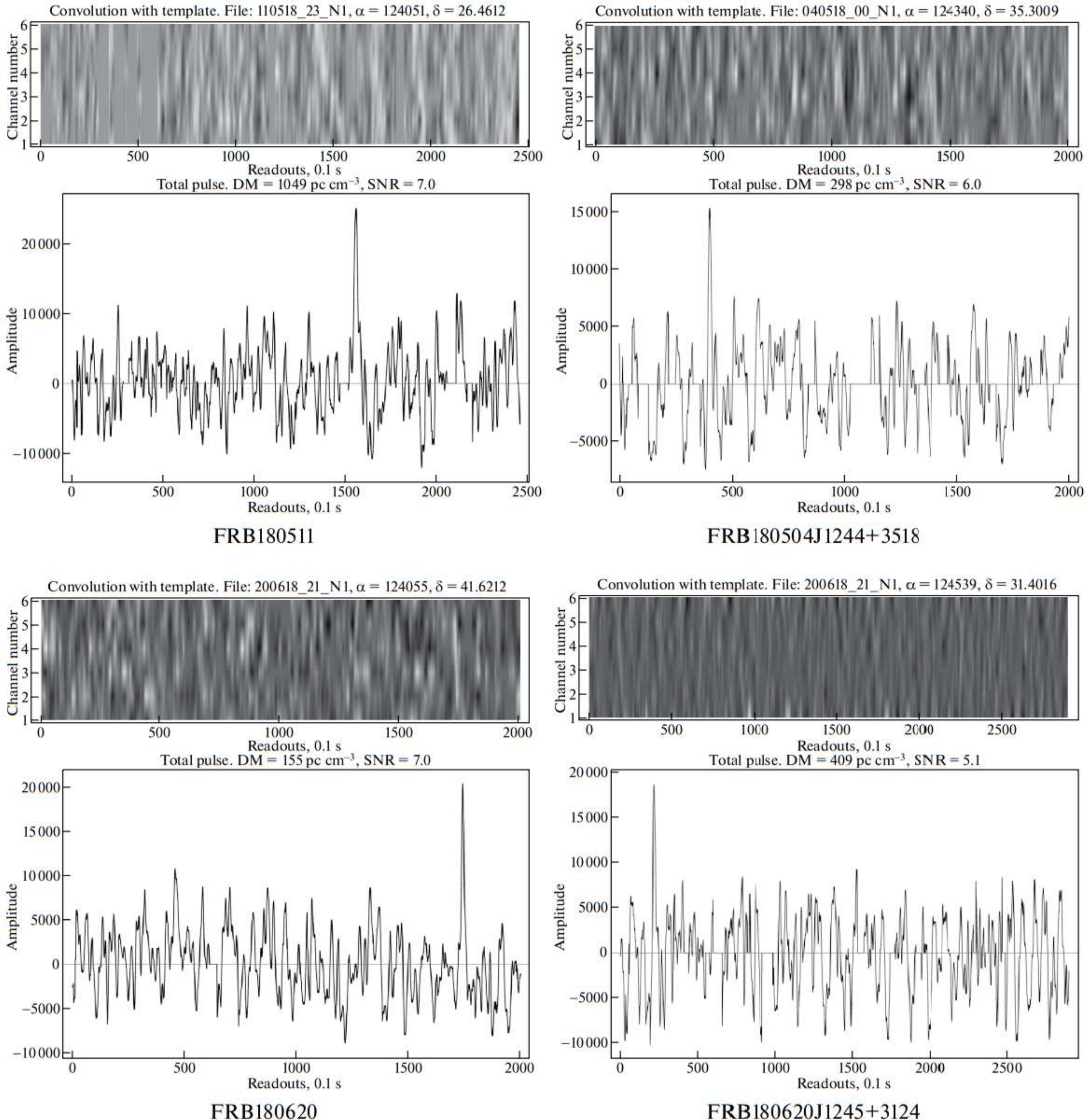}}
	\caption{Dynamic spectra and profiles of pulses.}
	\label{ris:fig24}
\end{figure}

\newpage
\begin{figure}[h!]
	\setcaptionmargin{1.3mm}
	\vbox{\includegraphics[width=1\linewidth]{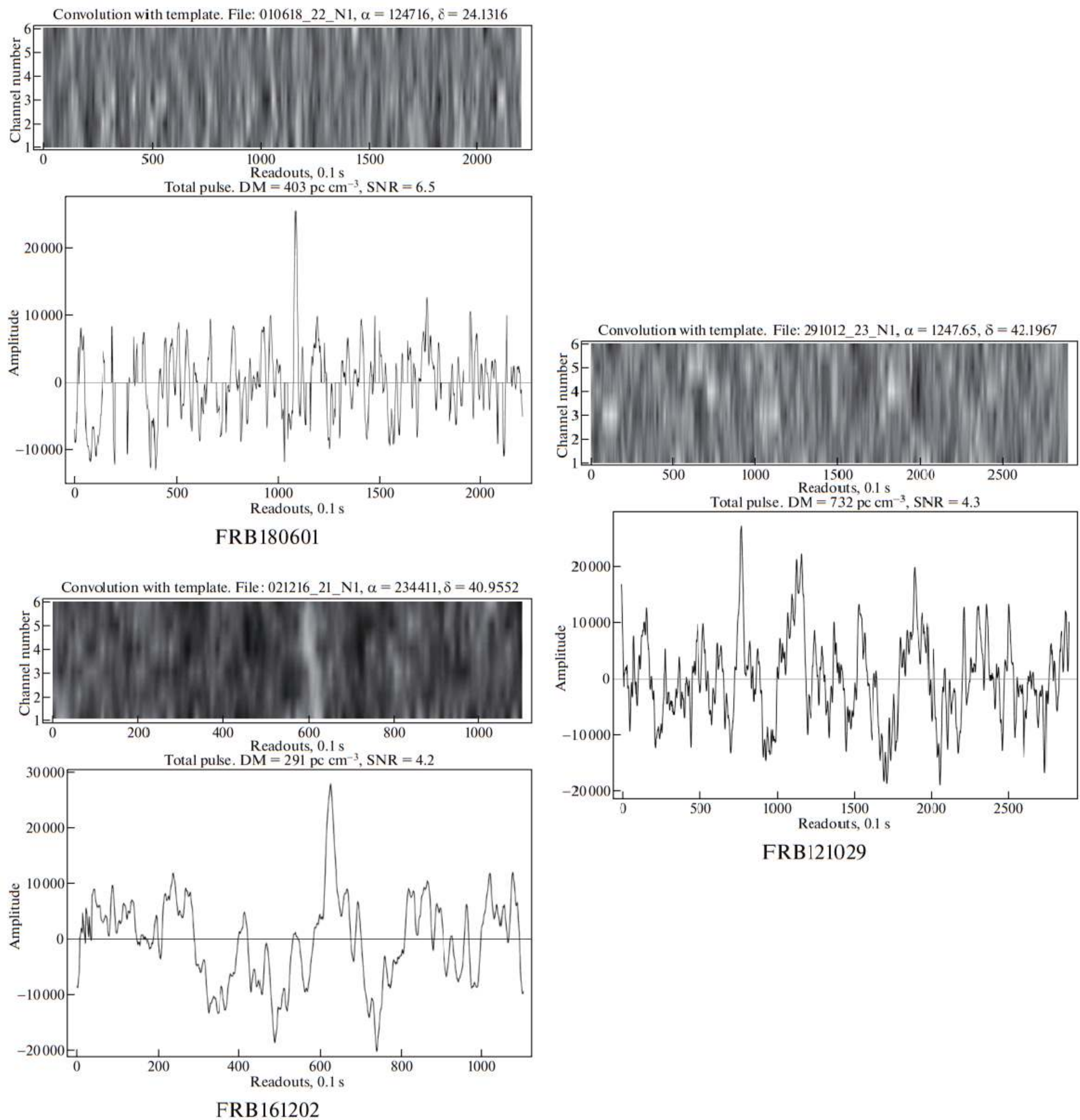}}
	\caption{Dynamic spectra and profiles of pulses.}
	\label{ris:fig25}
\end{figure}


\newpage
\section{REFERENCES}

\begin{thebibliography}{99}

\bibitem{Fuller} J. Fuller, Ch. D. Ott, MNRAS, {\bf 450}, 71 (2015).

\bibitem{Katz} J. I. Katz, MNRAS: Letters, {\bf 471}, 95 (2017).

\bibitem{Dai} Z. G. Dai1, J. S. Wang, X. F. Wu, Y. F. Huang, The Astrophysical Journal, {\bf 827(1)}, pp.7 (2016).

\bibitem{Margalit} B. Margalit, E. Berger, B. D. Metzger, ApJ, {\bf 886} (2019).

\bibitem{Istomin} Ya. N. Istomin, MNRAS, {\bf 478}, 4348 (2018).

\bibitem{Scholz} P. Scholz, CHIME/FRB Collaboration, The Astronomer's Telegram, No. 13681 (2020).

\bibitem{CHIME} CHIME/FRB Collaboration, M. Amiri, K. Bandura, et al., Nature, {\bf 566 (7743)}, 230 (2018).

\bibitem{ASKAP} K. W. Bannister, R. M. Shannon, J. P. Macquart et al., Astrophys Journal, {\bf 841}, L12 (2017).

\bibitem{Rodin} A. E. Rodin, V. V. Oreshko, V. A. Samodurov, Astronomy Reports, {\bf 61}, 30 (2017).

\bibitem{Fedorova1} V. A. Fedorova and A. E. Rodin, Astron. Rep., {\bf 63}, 39 (2019).

\bibitem{Fedorova2} V. A. Fedorova and A. E. Rodin, Astron. Rep., {\bf 63}, 877 (2019).

\bibitem{Cordes16} J. M. Cordes, R. S. Wharton, L. G. Spitler et al., arXiv:1605.05890v1 (2016).

\bibitem{Dolag} K. Dolag, B. M. Gaensler, A. M. Beck, V. C. Beck, MNRAS, {\bf 451}, 4277 (2015).

\bibitem{Scheuer} P. A. G. Scheuer, Nature, {\bf 218}, 920 (1968).

\bibitem{Rickett} B. J. Rickett, Nature, {\bf 221}, 158 (1969).

\bibitem{hewish} A. Hewish, S. J. Bell, J. D. H. Pilkington, P. F. Scott, R. A. Collins, Nature, {\bf 217 (5130)}, pp. 709 (1968).

\bibitem{Robinson} B. J. Robinson, B. F. C. Cooper, F. F. Gardiner, R. Wielebinski, T. L. Landecker, Nature, {\bf 218 (5147)}, pp. 1143 (1968).

\bibitem{Alekseev} Yu. I. Alekseev, V. V. Vitkevich, V. F. Zhuravlev, Yu. P. Shitov, Doklady Akademiia Nauk SSSR, Ser. Mat. Fiz., {\bf 187}, p. 1019 (1969).

\bibitem{Cordes} J. M. Cordes, J. M. Weisberg, V. Boriakoff, Astrophysical Journal, {\bf 288}, 221 (1985).

\bibitem {Sutton} M. J. Sutton, MNRAS, {\bf 155}, 51 (1971).

\bibitem {Bhat} N.D. Ramesh Bhat, James M. Cordes, F. Camilo et all, Astrophysical Journal, {\bf 605}, 759 (2004).

\bibitem {Lorimer} D. R. Lorimer, A. Karasteogiou, M. A. MacLaughlin, S. Jonson, MNRAS, {\bf 436}, L5 (2013).

\bibitem{Karastergiou} A. Karastergiou, J. Chennamangalam, W. Armour, et al. MNRAS, {\bf 452}, 1254 (2015).

\bibitem{Zhu} W. Zhu, L.--L. Feng, F. Zhang, The Astrophysical Jornal, {\bf 865 (2)}, 147 (2018).

\bibitem{Kuzmin2007} A. D. Kuz?min, B. Ya. Losovskii, and K. A. Lapaev,
Astron. Rep., {\bf 51}, 615 (2007).

\bibitem{Oppermann} N. Oppermann, L. D. Connor, U.-L. Pen, MNRAS {\bf 461}, 984 (2016).

\bibitem{Macquart} J. P. Macquart, R. D. Ekers, MNRAS, {\bf 474}, 1900 (2018).

\bibitem{Popov} S. B. Popov, K. A. Postnov, and M. S. Pshirkov, Phys.
Usp. {\bf 61}, 965 (2018).

\bibitem{Oreshko} V. V. Oreshko, G. A. Latyshev, I. A. Alekseev, et al.,
Tr. IPA {\bf 24}, 80 (2012).

\bibitem{frbcat} E. Petroff, L. Houben, K. Bannister, et al., Publications of the Astronomical Society of Australia, {\bf 33}, pp. 7 (2016).

\bibitem{PetroffBIG} E. Petroff, J. W. T. Hessels, D. R. Lorimer, The Astronomy and Astrophysics Review, {\bf 27}, 4 (2019).

\bibitem{YMW16} J. M. Yao, R. N. Manchester, N. Wang, Astrophysical Journal, {\bf 835(1)}, 32 (2017).

\bibitem{Zhu2} W. Zhu, L.-L.Feng, The Astrophysical Journal, {\bf 906}, p. 957 (2021).

\bibitem{CordLaz} J. M. Cordes, T. J. W. Lazio, arXiv:astro-ph/0301598 (2003).

\bibitem{Pilia} M. Pilia, M. Burgay, A. Possenti, et al., The Astrophysical Journal Letters, {\bf 896 (2)}, 11 (2020).

\bibitem{Sokolowski} M. Sokolowski, N. D. Bhat, J.P. Macquart, et al. The Astrophysical Journal Letters, {\bf 867}, L12 (2018).

\bibitem{Tingay}  S. J. Tingay, C. M. Trott, R. B. Wayth, et al., AJ, {\bf 150}, 199 (2015).

\bibitem{Chawla} P. Chawla, B. C. Andersen, M. Bhardwaj, M., et al., ApJL, {\bf 896}, L 41 (2020).

\bibitem{Houben} L. J. M. Houben,  L. G. Spitler, S. ter Veen, et al., AA, {\bf 623}, A42 (2019).

\end{thebibliography}
\end{document}